\documentclass[10pt,a4paper]{article}
\usepackage{graphicx}
\usepackage{array}
\usepackage[utf8]{inputenc}     
\usepackage[T1]{fontenc}   
\usepackage{mathptmx}      
\usepackage[left=35mm, right=35mm, top=15mm, bottom=20mm, noheadfoot]{geometry}

\usepackage{listings, color}
\usepackage{cite}
\usepackage{url}
\usepackage{amsfonts}
\usepackage{gnuplottex}

\begin{document}
\thispagestyle{empty}

\title{\textbf{Applying the Roofline Model for Deep Learning performance optimizations}}
\author{
Jacek Czaja \\ jacek.czaja@intel.com \\ \textsf{Intel Corporation}
\and
Michal Gallus \\ michal.gallus@intel.com \\ \textsf{Intel Corporation} 
\and
Adam Grygielski \\ adam.grygielski@intel.com \\ \textsf{Intel Corporation}
\and
Joanna Wozna \\ joanna.wozna@intel.com \\ \textsf{Intel Corporation}
\and
Tao Luo \\ luotao02@baidu.com \\ \textsf{Baidu}
}

\date{} 
\maketitle\thispagestyle{empty} 
\definecolor{mygray}{rgb}{0.8,0.8,0.8}
\begin{abstract}
In this paper We present a methodology for creating Roofline models automatically for Non-Unified Memory Access (NUMA*\cite{numa}) using Intel$^{\tiny{\textregistered}}$ Xeon as an example. Finally, we present an evaluation of highly efficient deep learning primitives as implemented in the Intel$^{\tiny{\textregistered}}$ (oneDNN) Library.
\end{abstract}

\section{Introduction}
Deep Learning is widely adopted for tasks, such as computer vision*\cite{alexnet} and natural language processing*\cite{DBLP:journals/corr/abs-1810-04805}. Deep learning tasks often require significant computational resources to operate on large datasets of labeled data. With those requirements in mind, existing hardware platforms are now being optimized for efficient deep learning execution.

One example is the development of the Intel® oneAPI Deep Neural Network (oneDNN) Library, which automatically implements operators, including convolution, matrix multiplication, pooling, batch normalization, activation functions, recurrent neural network (RNN) cells, and long short-term memory (LSTM) cells on x86 architectures, and accelerates inference performance using Intel Deep Learning Boost technology found on Intel Xeon Scalable processors. In this work we evaluated the Intel oneDNN library as on Intel Xeon processors using Roofline models.\\\\ 
The Roofline model is a methodology*\cite{roofline} for visual representation of platforms that can be used to:
\begin{itemize}
\item Estimate boundaries for performance gain from introducing new features e.g. multithreading and vectorization
\item Estimate limitations for improvement of a kernel's implementation
\item Explain efficiency of an existing kernel
\item Compare performance of computing platforms
\end{itemize}

\begin{figure}[h]
\input{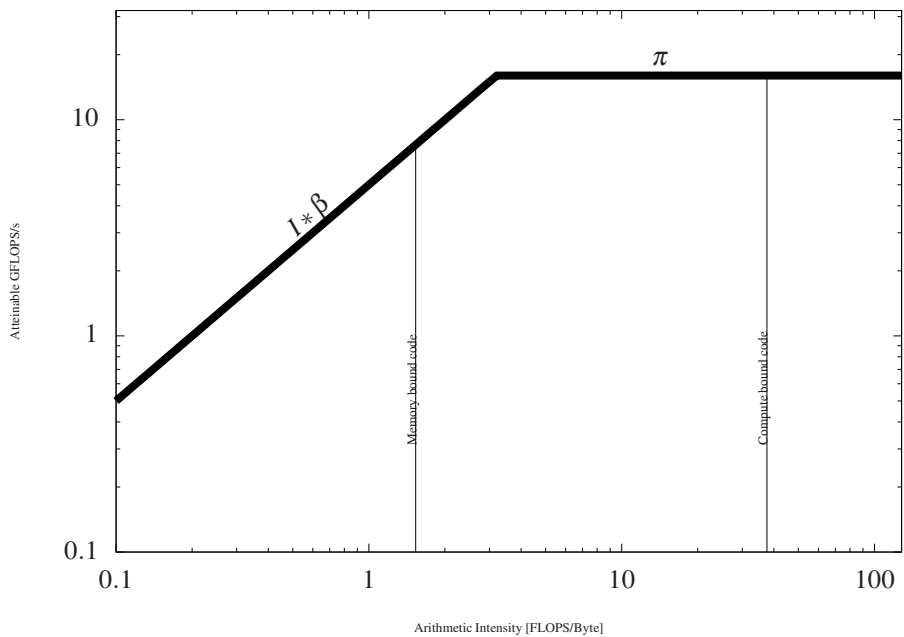}
\caption{simplified example of roofline}
\label{simple_roofline}
\end{figure}

The Roofline model ties a kernel's representation with platform capabilities (represented by roof), so evaluated kernel maximal performance is bounded by
the roof at a corresponding arithmetic intensity of kernel:
  \[
    P = \min \left\{
                \begin{array}{ll}
                  \pi\\
                  I*\beta
                \end{array}
              \right.
  \]
A simplified example is presented in Figure \ref{simple_roofline}.\\ 

This Roofline model relates the performance of the computer and memory traffic between the caches and DRAM. The model uses \textsf{arithmetic intensity}, (operations per byte of DRAM traffic), defining total bytes transferred to main memory after they have been filtered by the cache hierarchy. Thus, we obtained DRAM bandwidth needed by a kernel, what can discover bottleneck parts on the tested machine. 
The Roofline model is a 2D graph based on floating-point performance, memory performance and arithmetic intensity.\\\\ 

Initially, measurements needed for constructing the Roofline model were manually calculated. Offenbeck et al.*\cite{6844463} proposed a 
methodology for automatically obtaining needed measurements, based on Performance Monitoring units (PMU) of x86\_64 based computer architectures. This work is built on the above-mentioned article; we created a program to benchmark computing platforms and evaluate Deep Learning operators using a plot of the Roofline model for each evaluated platform and deep learning operator. We present our methodology on how to draw a Roofline model for Intel$^{\tiny{\textregistered}}$ Xeon processors with limited usage of resources: single core, single socket and two-sockets exection.\\\\
We chose to implement benchmarking code on our own both for better control over platform resources and educational purposes. We plotted roofline models for the oneDNN Library's deep learning primitives:
\begin{itemize}
\item activation (GELU) 
\item convolution
\item inner product
\item layer normalization
\item pooling (average)
\end{itemize}
 We then presented our observations from performed experiments.\\\\A conceptual description of the Roofline model is covered in detail in this article*\cite{roofline}. Practical meaning of \textsf{roofline} is that this shows (depending on the evaluated kernel's arithmetic intensity):
\begin{itemize}
\item Attainable compute utilization 
\item Possible gains from updating the kernel's implementation to use features like: multithreading or vectorization
\item Room for improvement of kernel's implementation for the same arithmetic intensity 
\end{itemize} 
To plot the Roofline model, we needed to gather characteristics of the computing platform and algorithm implementation (referred to as kernel) executed on that device, namely:
\begin{itemize}
\item Peak computational efficiency: \textsf{$\pi$}
\item Peak memory throughput: \textsf{T}
\item Amount of Floating point operations of kernel (Work) : \textsf{W}
\item Memory traffic of kernel: \textsf{Q}
\item Time of execution of kernel (Runtime): \textsf{R}
\end{itemize}

In section \ref{desc}, we describe how those attributes were measured in a context of our target CPU, the Intel Xeon Gold 6248 processor.

Once we agreed on methodology, we ran experiments on various oneDNN Library kernels; we present the results and observations of those activities in section \ref{analysis}

\section{Description of methodology} \label{desc}

All experiments were conducted on Intel Xeon Scalable processors with Intel \textsf{Turbo Boost} technology disabled, as suggested in the work*\cite{6844463} that we built upon.

\subsection{Measuring peak computational performance}
We chose to implement our own benchmark for checking peak computational capabilities so that we could better control resource usage for testing (threads, sockets). As well, we wanted our benchmarking to be independent from compiler optimizations, yet achieve maximum performance.
Our peak performance checking routine consists of independent execution of runtime-generated assembly code\ref{code1} on each of the available processor threads.  When implementing benchmark it is often a problem that compilers remove dead/unused code, but when benchmarking code is generated in runtime\footnote{using Xbyak*\cite{xbyak} project} we do not encounter that problem and overall performance is compiler-agnostic.  
\lstset{numbers=none}
\begin{figure}[ht]
\begin{lstlisting}[language=C,basicstyle=\small,escapechar=\%]
  ...
  vfmadd132ps zmm0,zmm1,zmm2
  vfmadd132ps zmm3,zmm1,zmm2
  vfmadd132ps zmm4,zmm1,zmm2
  vfmadd132ps zmm5,zmm1,zmm2
  vfmadd132ps zmm6,zmm1,zmm2
  vfmadd132ps zmm7,zmm1,zmm2
  ....
\end{lstlisting}
\caption{peak performance code snippet}
\label{code1}
\end{figure}

Assembly code is a sequence of FMA instructions from
Intel Advanced Vector Extensions (AVX, AVX2, AVX512) and designed to avoid chain dependency (read after write) so we can reach close to maximum performance. Using this benchmark we measured peak compute capabilities for the following processor scenarios:
\begin{itemize}
\item Single thread execution
\item Single socket execution
\item Two socket execution
\end{itemize}

\subsection{Measuring peak memory throughput} \label{peak_mem}
Measuring peak memory bandwidth is complicated, as results may vary depending on the operation we are measuring*\cite{McCalpin1995}. For that reason, we decided to determine maximal throughput value from independent checks:
\begin{itemize}
\item \textsf{memset} function from C standard library
\item \textsf{memcpy} functional from C standard library
\item Hand-crafted memset implemented in assembly language using non-temporal instructions 
\end{itemize}

Both types of benchmark were run single-threaded as well as multi-threaded and were processing .5 Gigabyte of memory. Our own implementation using non-temporal instructions was the fastest method when we ran experiments using for scenarios of single socket and two sockets. 
On the other hand, \textsf{memcpy} and \textsf{memset} reported higher memory throughput in the single-threaded scenario, which we attribute to the memory prefetching mechanism.\\\\We encountered an issue with our test of memory bandwidth in single-threaded situation as potentialy we could achieve higher bandwidth if we better utilized a memory prefetcher to benchmark memory bandwidth. This problem was present in some of Roofline plots with memory bound kernels. It may be that highly optimized, memory bound kernels executed in a single-threaded environment will have their actual runtime closer to (or beyond) the actual roof than necessary. For single socket or two socket based execution, this is not an important factor as the highest values of memory transferred are obtained using stream instructions (non-temporal stores).\\\\
One important thing to mention is that when running a bandwidth check on 
Intel Xeon processors, we bound memory allocations and threads of single-threaded and single-socket experiments to one chosen socket. It was needed as when full set of threads is used, there was not enough memory bandwidth 
available on one socket. We observed that threads and memory allocations were migrating to the second socket to take advantage of that socket's memory channels. This is generally efficient-wise practice, it was unwatned behaviour in our experiments as we wanted to limit execution to single socket.\\\\
Another important element is that to maximaize throughput when using two sockets (all available sockets of our target platform) we checked memory bandwidth by running two copies of our benchmarking program in parallel.
The threads and memory allocation of one running benchmark were bound to one node and second benchmark was bound to the second node. The sum of both throughputs was reported as the peak platform memory throughput.
Our justification for this merthod is that when threads are allocated on one node and memory is allocated on another node, it takes more time to access memory than when both resources are alloated on the same node.

\subsection{Counting Work}
Counting FLOPS was done in a similar way as described in this paper*\cite{6844463}.
The below, \textsf{perf} tool was used to read PMU counters:
\lstset{numbers=none}
\begin{lstlisting}[language=bash,basicstyle=\small,escapechar=\%]
FP_ARITH_INST_RETIRED:SCALAR_SINGLE
FP_ARITH_INST_RETIRED:128B_PACKED_SINGLE
FP_ARITH_INST_RETIRED:256B_PACKED_SINGLE
FP_ARITH_INST_RETIRED:512B_PACKED_SINGLE
\end{lstlisting}

We used \textsf{perf} externally to count \textsf{work} which made us to conduct two
measurements per evaluated kernel:
\begin{enumerate}
\item Run our testing program to perform single execution of kernel (overall counted)
\item Run our testing program to initialize all data, but do not perform
actual execution (framework overhead counted)
\end{enumerate}

Using PMU counter values from the above runs, we could subtract
framework overhead from overall measurement to get the value of counter for actual execution of the kernel. Next, we multiplied the counter value accordingly by 8 (for AVX2) and 16 (for AVX-512) to get actual FLOPS.\\ 

During this process, we had a concern if FLOPS were accuratly counted for FMA instructions, since the single FMA instruction for Intel AVX2 is actually performing 16 FLOPS, and for Intel AVX-512 it is performing 32 FLOPS. Therefore, we implemented the assembly code of \textsf{vfma132ps} (FMA instruction) and \textsf{vfaddps} (vector instruction of adding) and observed values of the PMU counter. We discovered that a single retirement of FMA instruction was increasing the counter by a factor of two as opposed to regular vector instructions where the counter was increased by one. This proved that FLOPS are counted precisely. As well, we implemented a more complex assembly code and compared its actuall FLOPS\footnote{Having code implemented in assembly made is easy to count executed FLOPS} with the FLOPS derived by te PMU counters-based method. Both results matched, so we concluded that this way of counting \textsf{work} is accurate.

\subsection{Counting memory traffic}
Determining memory traffic (Q) was the most challenging element of the
Roofline model to produce. Similar to work*\cite{6844463} we started by counting the 
memory transfer from last level cache to memory. This approach produced much lower
values than expected, due to memory prefetcher mechanisms. Next we disabled the hardware memory prefetcher as described here*\cite{disabling_prefetcher}. For simple evaluated kernels\footnote{For testing purposes of software solution created at that work we implemented sum reduction kernel} it provides accurate results, but for more complex algorithms like those implemented in Intel oneDNN Library, results were still much lower than expected. This is because the Intel oneDNN Library implementation is explicitly using software memory prefetcher instructions for GEMM and Winograd implementations which cannot be disabled by the methodology described in *\cite{disabling_prefetcher}.
Hence, we ended up in checking raw memory transfer as it goes through IMC (Internal memory controller) in a similar way as described in*\cite{6844463} . Since the modern Linux profiler \textsf{perf} was equipped*\cite{perf} with support of PMU counters of IMC, we did not have to add PMU counters on our own.\\\\ As IMCs' PMUs are counting memory transfer of the whole platform, not only CPU cores where execution of the evaluated algorithm takes place, counted traffic is not just related to the execution of the tested algorithm. Checking IMC uncore counters is available from command-line interface of perf, so to limit the measure of traffic only to the execution of our evaluated kernel, we
inspected the source code of the perf tool to get parameters values of syscall communicating perf with Linux kernel. With this knowledge we could call the same syscall in our code.\\\\ 
This method gives satisfying results for processed data greater than a megabyte. Analysis presented in this work is limited to algorithms that process bigger data (throughput) rather than a single chunk of data (latency).

\subsection{Measuring runtime performance}
We measured the time to conduct a number executions and reported an average value as \textsf{runtime}. We were interested in measuring
performance in three use cases:
\begin{itemize}
\item single-threaded execution
\item single-socket execution
\item two-sockets execution
\end{itemize}
We found that it was needed to control threads and memory allocations with \textsf{numactl} utility for the single-socket execution scenario. It proved to be a crucial element, as when all threads from the same socket
are heavily accessing memory then there is a shortage in memory bandwidth. The operating system may then migrate threads and allocation into another socket to use some of memory bandwidth of the other socket. This is the same situation
as described in section \ref{peak_mem}. Not having this restriction (e.g. controlling placement of resources with NUMA tools), will
result in a runtime performance that is higher than the actual roof for the analyzed kernel's arithmetic intensity.  

\subsubsection{Cold caches measurements}
We decided to clear caches for each iteration before measuring the 
execution time of the kernel. It was reported\cite{6844463} to invalidate measures when data size small, but for our experiments the buffer size was quite large \footnote{based on actually used sizes in Deep Learning workloads}, so
we did not see a problem with unstable measurements. The only problem was that overwriting caches is time consuming, which the running time our experiments.

\subsubsection{Warm caches measurements}
Before conducting actual measurements, we executed the actual kernel a number oftimes to have caches warmed and then performed the executions to be measured. Modern architectures have advanced
memory prefetching mechanisms built-in, so from that point of view the difference between cold and warm caches may not always be noticeable, in particular in some of oneDNN kernels that use software prefetching instructions.

\section{Analysis of Deep Learning Kernels}  \label{analysis}

\subsection{Analysis of Convolution}

In convolutional neural networks (CNNs), the majority of execution time is often spent
in the convolution operation itself. The Intel oneDNN Library provides efficient implementations of convolutions for various x86\_64 architectures. Roofline plots were generated by a program created for the purpose of this work. Our target processor for which we ran analysis in this work is the Intel Xeon Gold 6248 CPU. This processor has 44 cores, spread evenly between two sockets and is of NUMA architecture, as access time
to the same memory location from each core may differ. We ran analysis for three scenarios: 
\begin{itemize}
\item single threaded execution (Figure \ref{roof_conv_6248_single}) 
\item one socket execution (Figure \ref{roof_conv_6248_socket}) 
\item two socket execution (Figure \ref{roof_conv_6248_full})
\end{itemize}

\subsubsection{Single-threaded execution analysis}
\begin{figure}[h]
\input{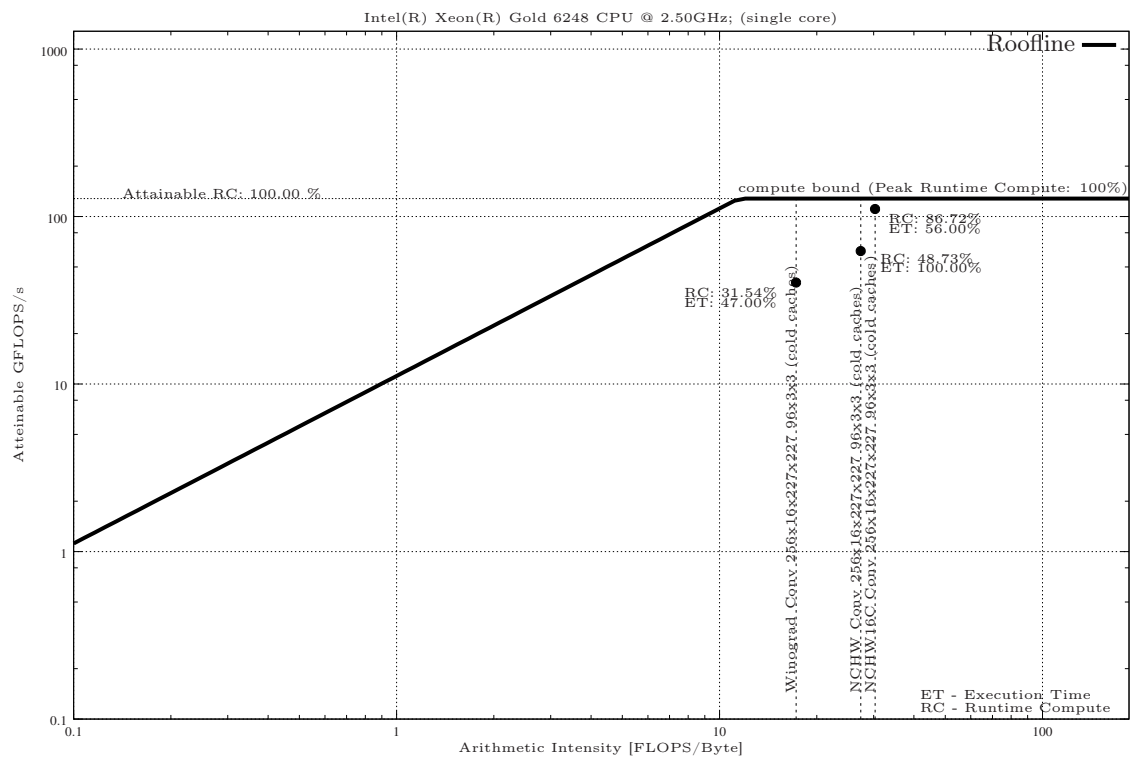}
\caption{Roofline plot with convolutional kernels using single thread of Intel Xeon Gold 6248 CPU}
\label{roof_conv_6248_single}
\end{figure}

We started our analysis of the convolution operation using only single-threaded execution. This is an applicable use case for the PaddlePaddle*\cite{paddlepaddle, Yanjun} deep learning framework
which is optimized for single-threaded execution. Figure \ref{roof_conv_6248_single} 
presents roofline plots\footnote{For the purpose of this article, absolute benchmarked values were turned into relative percentage measure}. We plotted the Roofline model of convolution operations using a fixed size of data to process in three sub-cases (vertical dashed lines from left to right in the Figure \ref{roof_conv_6248_socket}):
\begin{itemize}
\item Execution of convolution using Winograd*\cite{DBLP:journals/corr/Lavin15b} algorithm with cold caches
\item Execution of convolution using NCHW data arrangement with cold caches
\item Execution of convolution using NCHW16C (blocked) data arrangement with cold caches
\end{itemize}

First, we had three different convolutional kernels on the Roofline plot. Apart from
the relative utilization of compute capabilities (runtime compute) we also measured relative execution time (ET). NCHW convolution is the slowest so we denoted its ET as \textsf{100\%}. We can
see that the NCHW16C convolutional kernel is slightly more efficiently implemented as it utilizes \textsf{86\%} of peak compute, as opposed to the NCHW convolutional kernel which uses only \textsf{48\%} of available computational resources. This is quite intuitive; we compare two different implementations, conceptually the same kind of algorithm is performing same mathematical operations using roughly the same amount of FLOPS. Winograd convolution on the other hand, is a totally different algorithm, which ultimately produces the same results using a different calculation method. Hence, comparing kernels when implementing totally different algorithms has very limited sense. It is more on how well a given kernel will utilize computing platform resources. We can see that Winograd convolution utilization is much lower (\textsf{31\%}), yet it is the fastest one among the three presented.  
What we can see is that the implementation of Winograd has a room for improvement as its runtime compute is far from roof. Although Winograd is the fastest, its applications are limited to specific sizes of convolutional kernels , so direct convolution algorithm is of much wider use.\\\\Next we looked to compare two implementations of direct convolution \textsf{NCHW} versus cache and vectorization-friendly \textsf{NCHW16C}. The Intel oneDNN Library is implementing the idea of \textsf{layout propagation}*\cite{georganas2018anatomy} in a way that convolutional models input is converted from its original data arrangement to a blocked data arrangement (for example NCHW8C or NCHW16C). Then all subsequent deep learning operations (convolutions, normalization, non-linearities) work on this data arrangement. Blocked data arrangements help to ensure that all data used by vector instruction\footnote{AVX,AVX2, AVX512..} comes from the same single cacheline thus reducing memory latency and helping to achieve higher computational utilization.\\\\ We can see that the percentage of total compute utilization is much higher for \textsf{NCHW16C} than for \textsf{NCHW} data arrangement. Most compute friendly scenarios, such as convolution executed using \textsf{NCHW16C} data layout, achieve over \textsf{86.0\%} of maximal FLOPS available on the processor. Such a high compute utilization rate indicates that further optimization of this implementation (without conceptual redesigning or changing the convolutional algorithm) will be difficult. It may be easier to change algorithm to more efficient if one exists. One option may be to replace direct convolution with Winograd*\cite{DBLP:journals/corr/Lavin15b} convolution (if applicable) as discussed at the beginning of this section.

\subsubsection{Single socket execution analysis} \label{conv_socket}
In Figure \ref{roof_conv_6248_socket}, when comparing to single core execution (previous section), we can see that the respective compute resources utilization is slightly lower:
\begin{itemize}
\item Winograd convolution: from \textsf{31.54\%} to \textsf{29.30\%} 
\item Direct NCHW convolution: from \textsf{48.73\%} to \textsf{45.68\%} 
\item Direct NCHW16C convolution: from \textsf{86.72\%} to \textsf{78.01\%} 
\end{itemize}

\begin{figure}[h]
\input{conv_socket.tex}
\caption{Roofline plot with convolutional kernels using one-socket of Intel$^{\tiny{\textregistered}}$ Xeon$^{\tiny{\textregistered}}$ 6248}
\label{roof_conv_6248_socket}
\end{figure}

We attribute it partially to multi threads handling and partially to memory prefetcher / cache limitations. Without more deeper analysis it is difficult to draw a different conclusion other than that it is easier to implement an efficient single-threaded kernel than a multi-threaded one.\\\\
Another observation drawn from the presented Roofline model is that as we migrate execution of evaluated convolutions from a single thread to one socket or to two sockets execution, we can see that less efficient implementations are starting to become memory bound. The explanation for this is not related to the algorithms, it is that the rigid point of the Roofline model was moved further right.This is because memory bandwidth available per thread when using all hardware threads are available is lower than in the case of single thread execution. 

\subsubsection{Two socket execution analysis}
 As mentioned earlier, the Intel Xeon Gold 6248 has NUMA architecture. In this experiment we ran analysis on all available computing and memory resources to check utilization and compare it with single socket execution (subsection \ref{conv_socket})

\begin{figure}[h]
\input{conv_full.tex}
\caption{Roofline plot with convolutional kernels using two-sockets of Intel Xeon 6248}
\label{roof_conv_6248_full}
\end{figure}

Figure \ref{roof_conv_6248_full} presented our results using the full capabilities of the evaluated processor. We can see that the percentage of total compute utilization is relatively lower(\textbf{48\%}) to single socket execution (\textbf{78\%}) in cache friendly use case (\textsf{NCHW16C}) as well for the other two kernels’ executions. We checked that for both execution scenarios the same implementation is being executed, hence we are looking at how well the Intel oneDNN’s convolution execution scales from one socket to two sockets. Lower utilization of computing resources in a two-socket scenario is caused by the difficulty in harnessing the full computing resources of NUMA architecture with single kernel execution.\\\\ 

\subsection{Analysis of Inner Product}
\begin{figure}[h]
\input{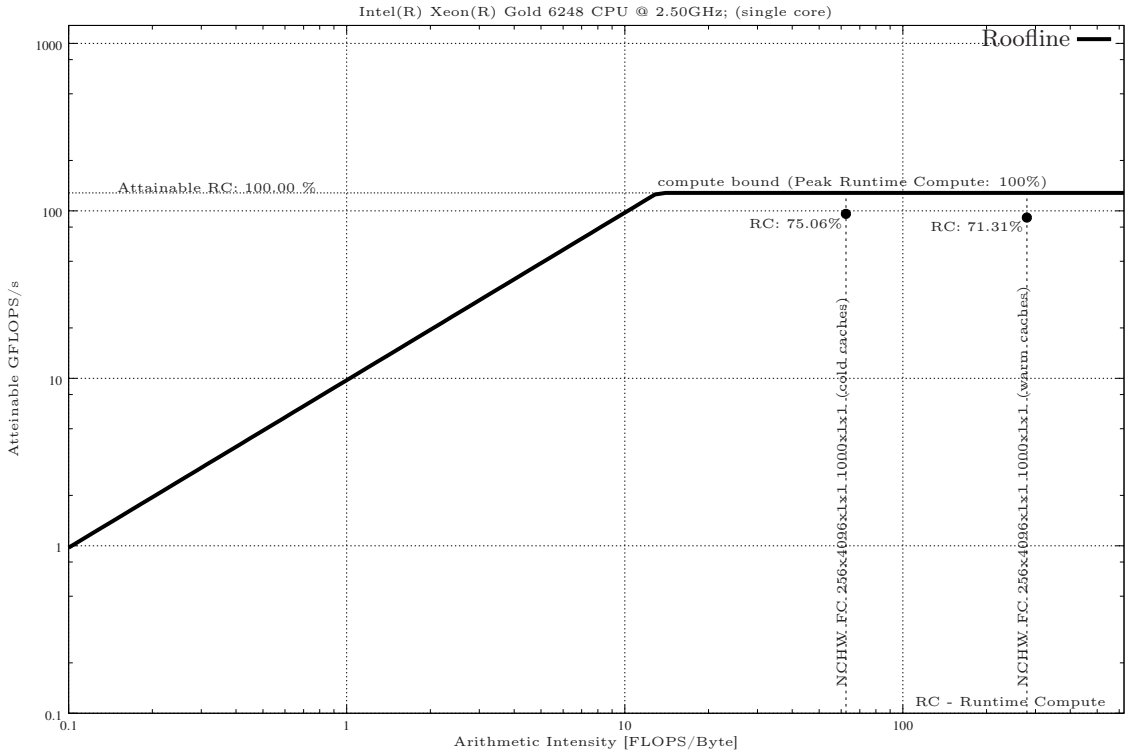}
\caption{Roofline plot of single-threaded Inner product}
\label{inner_prod_single}
\end{figure}

In this section we will look at Inner Product which is the base of neural networks . In particular, in modern natural language processing (NLP) solutions like transformer based models *\cite{DBLP:journals/corr/VaswaniSPUJGKP17},
the inner product takes majority of the execution time. The size of processed data of Inner Product as presented (Figure \ref{inner_prod_single}) does fit into the L3 cache of processor\footnote{Intel Xeon 6248} that was used. Hence it should be possible to observe a difference between execution with cold caches vs execution with caches warmed up.

Looking at the Roofline model, we can conclude that in the case of warmed caches, memory traffic is much smaller than when caches are cold. Although we execute the same code so the \textsf{work} is the same, the
arithmetic intensity is much higher in case of warmed caches execution, so memory traffic in that case has to be much smaller. Modern processors are using a memory prefetcher for reading data, which makes it difficult to predict memory traffic*\cite{article} \\\\
The other conclusion we can draw is that the Intel oneDNN Library's inner product is well optimized for that particular shape of input signal as runtime efficiency reaches over \textbf{71\%} of peak computational capacity for what is available on single-threaded execution. Roofline plots for other scenarios (e.g. execution using single socket and two-sockets) are in the appendix.

\subsection{Analysis of Pooling}
We attempted to analyze the pooling primitive using the Roofline model using two most popular pooling algorithms:
\begin{itemize}
\item max pooling 
\item average pooling 
\end{itemize}

For max pooling, the methodology used in this work is not applicable to this operation as max pooling consists of data movement and \textsf{max} operation which 
are not recognized as FLOPS and not traced by relevant FLOPS PMU counters. Therefore the \textsf{work} value will be counted will not be representative and useful. In this paper, we present only the Roofline plots for average pooling. 

\begin{figure}[h]
\input{pool_single.tex}
\caption{Roofline plot with Average Pool kernels single-threaded Intel$^{\tiny{\textregistered}}$ Xeon$^{\tiny{\textregistered}}$ 6248}
\label{roof_pool_6248_single}
\end{figure}

Figure \ref{roof_pool_6248_single} shows that arithmetic intensity for \textsf{NCHW} and blocked layout data arrangement (\textsf{NCHW16C}) in a situation with cold caches is almost the same. The same observation
applies to the warmed caches scenario. This is not very surprising in itself, but an interesting observation is that there is a huge difference in the percentage of CPU compute utilization. Implementations using \textsf{NCHW} 
data arrangement achieved \textbf{0.35\%} of compute utilization and \textsf{NCHW16C} implementation are utilizing around \textbf{~14.8 \%} which is over 42 x better utilization. We found this interesting
and searched for an explanation.\\\\
The Intel oneDNN library can work in \textsf{verbose} mode to provide details of internal execution as presented below:
\begin{itemize}
\item \textsf{NCHW:}
\lstset{numbers=none}
\begin{lstlisting}[language=bash,basicstyle=\small,escapechar=\%]
dnnl_verbose,exec,cpu,pooling,simple_nchw:any,forward_inference,...
\end{lstlisting}
\item \textsf{NCHW16C:}
\lstset{numbers=none}
\begin{lstlisting}[language=bash,basicstyle=\small,escapechar=\%]
dnnl_verbose,exec,cpu,pooling,jit:avx512_common,forward_inference,...
\end{lstlisting}
\end{itemize}

Based on those outputs we can see that \textsf{NCHW} is using an average pooling implementation named : \textsf{simple\_nchw} and the blocked data arrangement is using \textsf{jit::avx512\_common} implementation. The former is a \textsf{C++} based naive implementation and the latter one is a runtime generated assembly code that was implemented using the Xbyak*\cite{xbyak} project. NCHW pooling requires doing operations with-in simd register (as spatial has stride 1), while NHWC and NCHW16C pooling could directly operate on registers. This is the primary reason for NCHW being that low on compute utilization.  

\subsection{Analysis of GELU activation}
Another oneDNN primitive we analyzed was recently introduced into oneDNN Gaussian Error Linear Units*\cite{DBLP:journals/corr/HendrycksG16} (GELU) activation.
The reason why we chose to analyze that one is that GELU is an element-wise operation so data arrangement should not have an impact on performance of execution. Moreover activations are of lesser arithmetic intensity compared to convolutions as they are memory bound and we wanted to check if our work is applicable to memory bound primitives as well. 

\begin{figure}[h]
\input{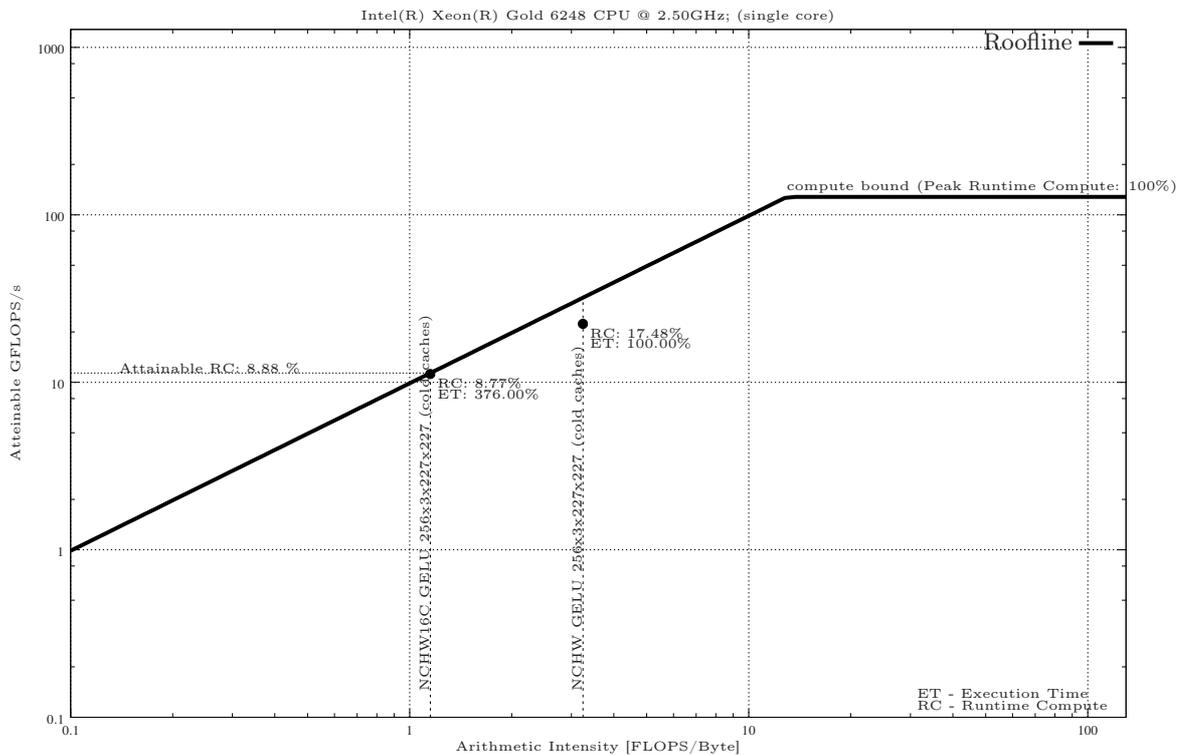}
\caption{Roofline plot with GELU kernels using single-core of Intel Xeon Gold 6248}
\label{roof_gelu_6248_single}
\end{figure}

The presented roofline model (Figure \ref{roof_gelu_6248_single}) shows GELU operation 
executed via Intel oneDNN library. To our surprise We observed that execution using block data arrangement (NCHW16C) is of lower arithmetic intensity that NCHW implementation. We expected the same performance on both data arrangements. But when looking into actual values of \textsf{work} and \textsf{traffic} we saw that NCHW16C consumed four times as much memory and twice as much FLOPS than NCHW implementation. Our roofline model plots as such does not show W and T values and seeing lower Operation Intensity for NCHW16C made as to check underneath gathered data.\\\\ Explanation of why twice as much of resources is consumed is that dimensionality of input
 signal [256,3,227,227] is having a second dimension (channel) equal to 3. Efficient implementation of NCHW16C, NCHW8C as provided by oneDNN, require that channel value is 
multiplication of 8. Hence oneDNN when \textbf{forced} to use blocked data arrangement and as a consequence to extend input tensor to have a shape of [256,8,227,227]. In that situation it is less efficient that using NCHW data arrangement. Does it mean that user has to understand details of implementation of oneDNN kernels to use them efficiently? The answer is No as Intel oneDNN library usage model is that computational primitives are choosing on their own which implementation to use. For the purpose of analysis we chose both unfavorable dimensionality for
blocked processing and enforced GELU to work on blocked processing. In other words oneDNN internal logic will trigger most efficient implementation , which at that situation 
GELU working on blocked layout is not.  GELU roofline plots far more often encountered dimensionality (more favorable to oneDNN) as presented in Appendix are confirming that 
arithmetic intensity for blocked and NCHW data arrangements are very similar as well over all efficiency.

\subsection{Applicability of methodology}
Work(W) is measured via FLOPS PMU events which counts floating point operations like subtractions, additions and multiplications, but it does not
count data movements as well as getting maximum values which may be implemented using vmaxps instructions of AVX instruction set. This implies that measuring Work using PMU events (as used throughout this article) is not suitable methodology when to analyze Deep Learning algorithms like Rectified Linear Units (ReLU), Max-pooling and reorders and other primitives where
majority of work is performed by operations not considered when counting FLOPS e.g. \textsf{min}, \textsf{max} and data movements. 

\section{Conclusion and Further work}

During this work we found Roofline models to be an effective tool to help in understanding a complex deep learning library as Intel oneDNN Library, both from
user perspective and oneDNN developer's as well. We expected that this project we developed may be very helpful in validation of oneDNN's kernels\\\\
A natural extension of this work would be applying roofline model to integer based computations
as it is supported by oneDNN and widely used in modern Deep learning workloads*\cite{int8-2}. It would be also good to evaluate others operation like: \textsf{max} and \textsf{min} and data movements to have a full evaluation of oneDNN's operations. Both mentioned directions depend on the availability of existence of relevant PMU events.\\\\
Another direction to consider would be to improve the methodology of creating the Roofline models for single core scenarios. In this work for single core roofline we just benchmarked peak computation and bandwidth as available using single thread. It is fine for peak computation as this will scale with number of cores used. But situation is different for checking bandwidth, due to memory prefetcher
working in background. Regardless of the number of cores of single socket used we always have the same number of memory prefetcher streams. So Memory bandwidth will not scale linearly as we increase number of cores used. Perhaps it would make more sense to run benchmarking in parallel for each core independently and report the average value as a bandwidth available for a single core. 

\section*{Acknowledgment}
The authors would like to express our gratitude to Krzysztof Badziak and Mateusz Ozga from Intel Corporation for their advice on optimizations and to Andres Rodriguez, Evarist Fomenko and Emily Hutson for reviewing this article and to Michal Lukaszewski and Michal Chruscinski for providing  and preparing platform to run experiments on. 

\bibliography{articles}{}
\bibliographystyle{plain}

\section*{Appendix: Notices and Disclaimers}

Software and workloads used in performance tests may have been optimized for performance only on Intel microprocessors. 

Performance tests, such as SYSmark and MobileMark, are measured using specific computer systems, components, software, operations and functions. Any change to any of those factors may cause the results to vary. You should consult other information and performance tests to assist you in fully evaluating your contemplated purchases, including the performance of that product when combined with other products.   For more complete information visit:\\\\ \url{www.intel.com/benchmarks}.  
\subsubsection*{Configurations:} \label{configurations} 

Project we developed as part of this is work is currently publicly not-available.\\
We evaluated oneDNN project version tagged as: GIT\_TAG v1.2, publicly available at:\\\url{https://github.com/intel/mkl-dnn}.\\\\
\vskip 0.05cm
All measures and performance evaluation as presented in this article were taken using Intel$^{\tiny{\textregistered}}$ Xeon$^{\tiny{\textregistered}} $ 6248 processor running \textsf{Ubuntu 18.04 Linux}\\\\
Performance results are based on testing as of 8th of July 2020 and may not reflect all publicly available security updates.  No product or component can be absolutely secure.\\\\
Optimization Notice: Intel's compilers may or may not optimize to the same degree for non-Intel microprocessors for optimizations that are not unique to Intel microprocessors. These optimizations include SSE2, SSE3, and SSSE3 instruction sets and other optimizations. Intel does not guarantee the availability, functionality, or effectiveness of any optimization on microprocessors not manufactured by Intel. Microprocessor-dependent optimizations in this product are intended for use with Intel microprocessors. Certain optimizations not specific to Intel microarchitecture are reserved for Intel microprocessors. Please refer to the applicable product User and Reference Guides for more information regarding the specific instruction sets covered by this notice.\\\\
Notice Revision \#20110804\\\\
Intel$^{\tiny{\textregistered}}$, the Intel$^{\tiny{\textregistered}}$ logo, and Intel$^{\tiny{\textregistered}}$ Xeon$^{\tiny{\textregistered}}$ are trademarks of Intel Corporation or its subsidiaries in the U.S. and/or other countries.\\\\
\textcopyright Intel Corporation

\section*{Appendix: Roofline plots of oneDNN kernels}

\subsection*{Layer Normalization}

\begingroup
  \fontfamily{Helvetica}%
  \selectfont
  \makeatletter
  \providecommand\color[2][]{%
    \GenericError{(gnuplot) \space\space\space\@spaces}{%
      Package color not loaded in conjunction with
      terminal option `colourtext'%
    }{See the gnuplot documentation for explanation.%
    }{Either use 'blacktext' in gnuplot or load the package
      color.sty in LaTeX.}%
    \renewcommand\color[2][]{}%
  }%
  \providecommand\includegraphics[2][]{%
    \GenericError{(gnuplot) \space\space\space\@spaces}{%
      Package graphicx or graphics not loaded%
    }{See the gnuplot documentation for explanation.%
    }{The gnuplot epslatex terminal needs graphicx.sty or graphics.sty.}%
    \renewcommand\includegraphics[2][]{}%
  }%
  \providecommand\rotatebox[2]{#2}%
  \@ifundefined{ifGPcolor}{%
    \newif\ifGPcolor
    \GPcolorfalse
  }{}%
  \@ifundefined{ifGPblacktext}{%
    \newif\ifGPblacktext
    \GPblacktexttrue
  }{}%
  \let\gplgaddtomacro\g@addto@macro
  \gdef\gplbacktext{}%
  \gdef\gplfronttext{}%
  \makeatother
  \ifGPblacktext
    \def\colorrgb#1{}%
    \def\colorgray#1{}%
  \else
    \ifGPcolor
      \def\colorrgb#1{\color[rgb]{#1}}%
      \def\colorgray#1{\color[gray]{#1}}%
      \expandafter\def\csname LTw\endcsname{\color{white}}%
      \expandafter\def\csname LTb\endcsname{\color{black}}%
      \expandafter\def\csname LTa\endcsname{\color{black}}%
      \expandafter\def\csname LT0\endcsname{\color[rgb]{1,0,0}}%
      \expandafter\def\csname LT1\endcsname{\color[rgb]{0,1,0}}%
      \expandafter\def\csname LT2\endcsname{\color[rgb]{0,0,1}}%
      \expandafter\def\csname LT3\endcsname{\color[rgb]{1,0,1}}%
      \expandafter\def\csname LT4\endcsname{\color[rgb]{0,1,1}}%
      \expandafter\def\csname LT5\endcsname{\color[rgb]{1,1,0}}%
      \expandafter\def\csname LT6\endcsname{\color[rgb]{0,0,0}}%
      \expandafter\def\csname LT7\endcsname{\color[rgb]{1,0.3,0}}%
      \expandafter\def\csname LT8\endcsname{\color[rgb]{0.5,0.5,0.5}}%
    \else
      \def\colorrgb#1{\color{black}}%
      \def\colorgray#1{\color[gray]{#1}}%
      \expandafter\def\csname LTw\endcsname{\color{white}}%
      \expandafter\def\csname LTb\endcsname{\color{black}}%
      \expandafter\def\csname LTa\endcsname{\color{black}}%
      \expandafter\def\csname LT0\endcsname{\color{black}}%
      \expandafter\def\csname LT1\endcsname{\color{black}}%
      \expandafter\def\csname LT2\endcsname{\color{black}}%
      \expandafter\def\csname LT3\endcsname{\color{black}}%
      \expandafter\def\csname LT4\endcsname{\color{black}}%
      \expandafter\def\csname LT5\endcsname{\color{black}}%
      \expandafter\def\csname LT6\endcsname{\color{black}}%
      \expandafter\def\csname LT7\endcsname{\color{black}}%
      \expandafter\def\csname LT8\endcsname{\color{black}}%
    \fi
  \fi
    \setlength{\unitlength}{0.0500bp}%
    \ifx\gptboxheight\undefined%
      \newlength{\gptboxheight}%
      \newlength{\gptboxwidth}%
      \newsavebox{\gptboxtext}%
    \fi%
    \setlength{\fboxrule}{0.5pt}%
    \setlength{\fboxsep}{1pt}%
\begin{picture}(8502.00,5668.00)%
    \gplgaddtomacro\gplbacktext{%
      \csname LTb\endcsname%
      \put(488,256){\makebox(0,0)[r]{\strut{}\tiny $0.1$}}%
      \csname LTb\endcsname%
      \put(488,1212){\makebox(0,0)[r]{\strut{}\tiny $1$}}%
      \csname LTb\endcsname%
      \put(488,2168){\makebox(0,0)[r]{\strut{}\tiny $10$}}%
      \csname LTb\endcsname%
      \put(488,3125){\makebox(0,0)[r]{\strut{}\tiny $100$}}%
      \csname LTb\endcsname%
      \put(488,4081){\makebox(0,0)[r]{\strut{}\tiny $1000$}}%
      \csname LTb\endcsname%
      \put(488,5037){\makebox(0,0)[r]{\strut{}\tiny $10000$}}%
      \csname LTb\endcsname%
      \put(536,176){\makebox(0,0){\strut{}\tiny $0.1$}}%
      \csname LTb\endcsname%
      \put(2814,176){\makebox(0,0){\strut{}\tiny $1$}}%
      \csname LTb\endcsname%
      \put(5092,176){\makebox(0,0){\strut{}\tiny $10$}}%
      \csname LTb\endcsname%
      \put(7370,176){\makebox(0,0){\strut{}\tiny $100$}}%
      \colorrgb{0.00,0.00,0.00}%
      \put(6079,4546){\makebox(0,0)[l]{\strut{}\tiny compute bound (Peak Runtime Compute: 100\%)}}%
      \put(2815,453){\rotatebox{90}{\makebox(0,0)[l]{\strut{}\tiny  TNC Layer Norm 128x256x768 (cold caches)}}}%
      \put(2180,2804){\makebox(0,0)[l]{\strut{}\tiny RC: 2.32\%}}%
      \csname LTb\endcsname%
      \put(850,3162){\makebox(0,0)[l]{\strut{}\tiny Attainable RC: 3.89 \%}}%
      \colorrgb{0.00,0.00,0.00}%
      \put(2961,453){\rotatebox{90}{\makebox(0,0)[l]{\strut{}\tiny TNC Layer Norm 128x256x768 (warm caches)}}}%
      \put(3060,2888){\makebox(0,0)[l]{\strut{}\tiny RC: 2.46\%}}%
      \put(6801,357){\makebox(0,0)[l]{\strut{}\tiny{RC - Runtime Compute}}}%
    }%
    \gplgaddtomacro\gplfronttext{%
      \csname LTb\endcsname%
      \put(64,2841){\rotatebox{-270}{\makebox(0,0){\strut{}\tiny Atteinable GFLOPS/s}}}%
      \put(4446,56){\makebox(0,0){\strut{}\tiny Arithmetic Intensity [FLOPS/Byte]}}%
      \put(4446,5547){\makebox(0,0){\strut{}\tiny Intel(R) Xeon(R) Gold 6248 CPU @ 2.50GHz; (single socket)}}%
      \csname LTb\endcsname%
      \put(7958,5324){\makebox(0,0)[r]{\strut{}\small Roofline}}%
    }%
    \gplbacktext
    \put(0,0){\includegraphics{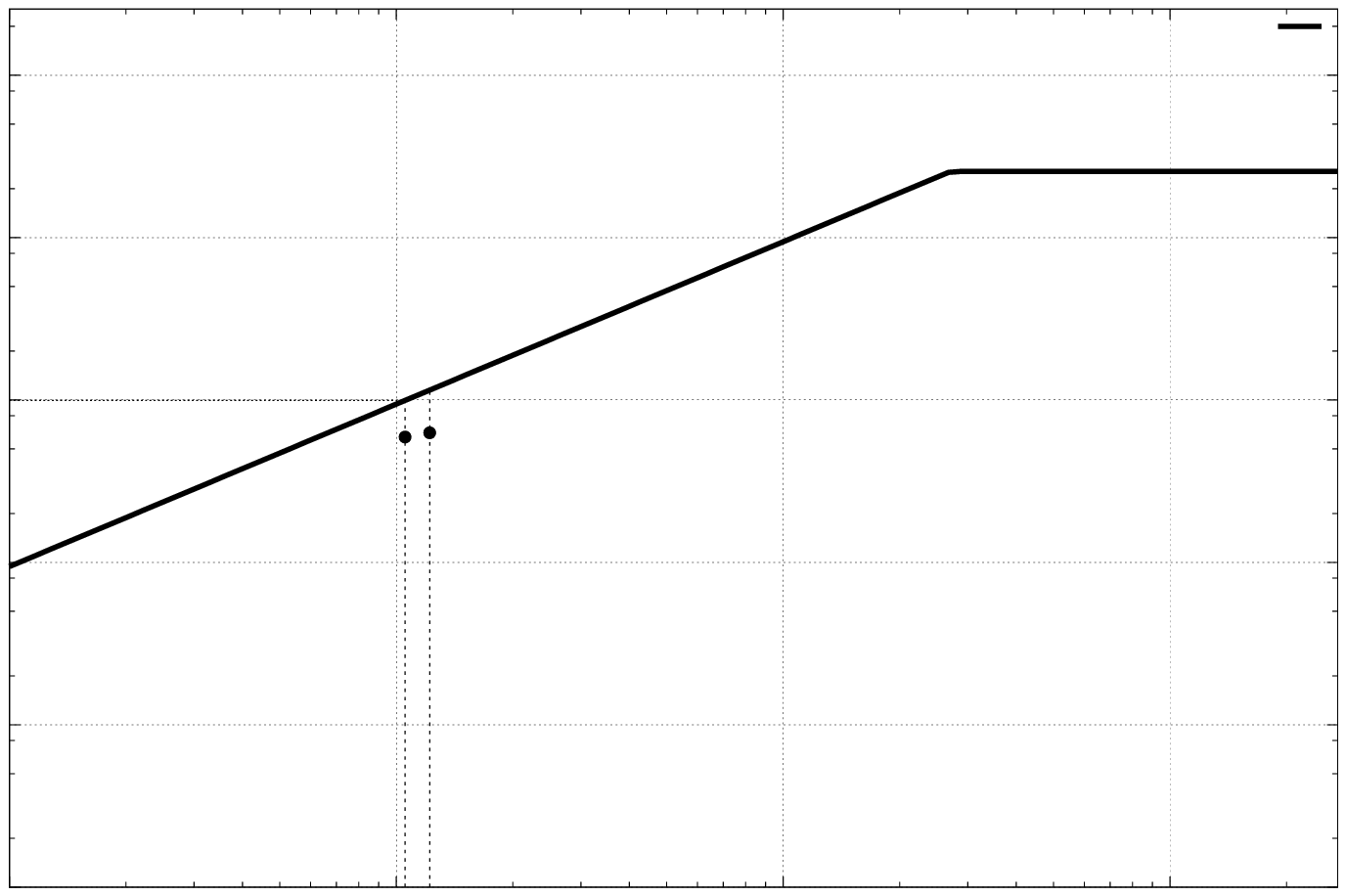}}%
    \gplfronttext
  \end{picture}%
\endgroup

\begingroup
  \fontfamily{Helvetica}%
  \selectfont
  \makeatletter
  \providecommand\color[2][]{%
    \GenericError{(gnuplot) \space\space\space\@spaces}{%
      Package color not loaded in conjunction with
      terminal option `colourtext'%
    }{See the gnuplot documentation for explanation.%
    }{Either use 'blacktext' in gnuplot or load the package
      color.sty in LaTeX.}%
    \renewcommand\color[2][]{}%
  }%
  \providecommand\includegraphics[2][]{%
    \GenericError{(gnuplot) \space\space\space\@spaces}{%
      Package graphicx or graphics not loaded%
    }{See the gnuplot documentation for explanation.%
    }{The gnuplot epslatex terminal needs graphicx.sty or graphics.sty.}%
    \renewcommand\includegraphics[2][]{}%
  }%
  \providecommand\rotatebox[2]{#2}%
  \@ifundefined{ifGPcolor}{%
    \newif\ifGPcolor
    \GPcolorfalse
  }{}%
  \@ifundefined{ifGPblacktext}{%
    \newif\ifGPblacktext
    \GPblacktexttrue
  }{}%
  \let\gplgaddtomacro\g@addto@macro
  \gdef\gplbacktext{}%
  \gdef\gplfronttext{}%
  \makeatother
  \ifGPblacktext
    \def\colorrgb#1{}%
    \def\colorgray#1{}%
  \else
    \ifGPcolor
      \def\colorrgb#1{\color[rgb]{#1}}%
      \def\colorgray#1{\color[gray]{#1}}%
      \expandafter\def\csname LTw\endcsname{\color{white}}%
      \expandafter\def\csname LTb\endcsname{\color{black}}%
      \expandafter\def\csname LTa\endcsname{\color{black}}%
      \expandafter\def\csname LT0\endcsname{\color[rgb]{1,0,0}}%
      \expandafter\def\csname LT1\endcsname{\color[rgb]{0,1,0}}%
      \expandafter\def\csname LT2\endcsname{\color[rgb]{0,0,1}}%
      \expandafter\def\csname LT3\endcsname{\color[rgb]{1,0,1}}%
      \expandafter\def\csname LT4\endcsname{\color[rgb]{0,1,1}}%
      \expandafter\def\csname LT5\endcsname{\color[rgb]{1,1,0}}%
      \expandafter\def\csname LT6\endcsname{\color[rgb]{0,0,0}}%
      \expandafter\def\csname LT7\endcsname{\color[rgb]{1,0.3,0}}%
      \expandafter\def\csname LT8\endcsname{\color[rgb]{0.5,0.5,0.5}}%
    \else
      \def\colorrgb#1{\color{black}}%
      \def\colorgray#1{\color[gray]{#1}}%
      \expandafter\def\csname LTw\endcsname{\color{white}}%
      \expandafter\def\csname LTb\endcsname{\color{black}}%
      \expandafter\def\csname LTa\endcsname{\color{black}}%
      \expandafter\def\csname LT0\endcsname{\color{black}}%
      \expandafter\def\csname LT1\endcsname{\color{black}}%
      \expandafter\def\csname LT2\endcsname{\color{black}}%
      \expandafter\def\csname LT3\endcsname{\color{black}}%
      \expandafter\def\csname LT4\endcsname{\color{black}}%
      \expandafter\def\csname LT5\endcsname{\color{black}}%
      \expandafter\def\csname LT6\endcsname{\color{black}}%
      \expandafter\def\csname LT7\endcsname{\color{black}}%
      \expandafter\def\csname LT8\endcsname{\color{black}}%
    \fi
  \fi
    \setlength{\unitlength}{0.0500bp}%
    \ifx\gptboxheight\undefined%
      \newlength{\gptboxheight}%
      \newlength{\gptboxwidth}%
      \newsavebox{\gptboxtext}%
    \fi%
    \setlength{\fboxrule}{0.5pt}%
    \setlength{\fboxsep}{1pt}%
\begin{picture}(8502.00,5668.00)%
    \gplgaddtomacro\gplbacktext{%
      \csname LTb\endcsname%
      \put(488,256){\makebox(0,0)[r]{\strut{}\tiny $0.1$}}%
      \csname LTb\endcsname%
      \put(488,1165){\makebox(0,0)[r]{\strut{}\tiny $1$}}%
      \csname LTb\endcsname%
      \put(488,2074){\makebox(0,0)[r]{\strut{}\tiny $10$}}%
      \csname LTb\endcsname%
      \put(488,2984){\makebox(0,0)[r]{\strut{}\tiny $100$}}%
      \csname LTb\endcsname%
      \put(488,3893){\makebox(0,0)[r]{\strut{}\tiny $1000$}}%
      \csname LTb\endcsname%
      \put(488,4802){\makebox(0,0)[r]{\strut{}\tiny $10000$}}%
      \csname LTb\endcsname%
      \put(536,176){\makebox(0,0){\strut{}\tiny $0.1$}}%
      \csname LTb\endcsname%
      \put(2828,176){\makebox(0,0){\strut{}\tiny $1$}}%
      \csname LTb\endcsname%
      \put(5120,176){\makebox(0,0){\strut{}\tiny $10$}}%
      \csname LTb\endcsname%
      \put(7411,176){\makebox(0,0){\strut{}\tiny $100$}}%
      \colorrgb{0.00,0.00,0.00}%
      \put(5843,4590){\makebox(0,0)[l]{\strut{}\tiny compute bound (Peak Runtime Compute: 100\%)}}%
      \put(2828,453){\rotatebox{90}{\makebox(0,0)[l]{\strut{}\tiny  TNC Layer Norm 128x256x768 (cold caches)}}}%
      \put(2189,2862){\makebox(0,0)[l]{\strut{}\tiny RC: 1.68\%}}%
      \csname LTb\endcsname%
      \put(850,3291){\makebox(0,0)[l]{\strut{}\tiny Attainable RC: 4.07 \%}}%
      \colorrgb{0.00,0.00,0.00}%
      \put(3005,453){\rotatebox{90}{\makebox(0,0)[l]{\strut{}\tiny  TNC Layer Norm 128x256x768 (warm caches)}}}%
      \put(3105,3046){\makebox(0,0)[l]{\strut{}\tiny RC: 2.67\%}}%
      \put(6801,357){\makebox(0,0)[l]{\strut{}\tiny RC - Runtime Compute}}%
    }%
    \gplgaddtomacro\gplfronttext{%
      \csname LTb\endcsname%
      \put(64,2841){\rotatebox{-270}{\makebox(0,0){\strut{}\tiny Atteinable GFLOPS/s}}}%
      \put(4446,56){\makebox(0,0){\strut{}\tiny Arithmetic Intensity [FLOPS/Byte]}}%
      \put(4446,5547){\makebox(0,0){\strut{}\tiny Intel(R) Xeon(R) Gold 6248 CPU @ 2.50GHz; (two sockets)}}%
      \csname LTb\endcsname%
      \put(7958,5324){\makebox(0,0)[r]{\strut{}\small Roofline}}%
    }%
    \gplbacktext
    \put(0,0){\includegraphics{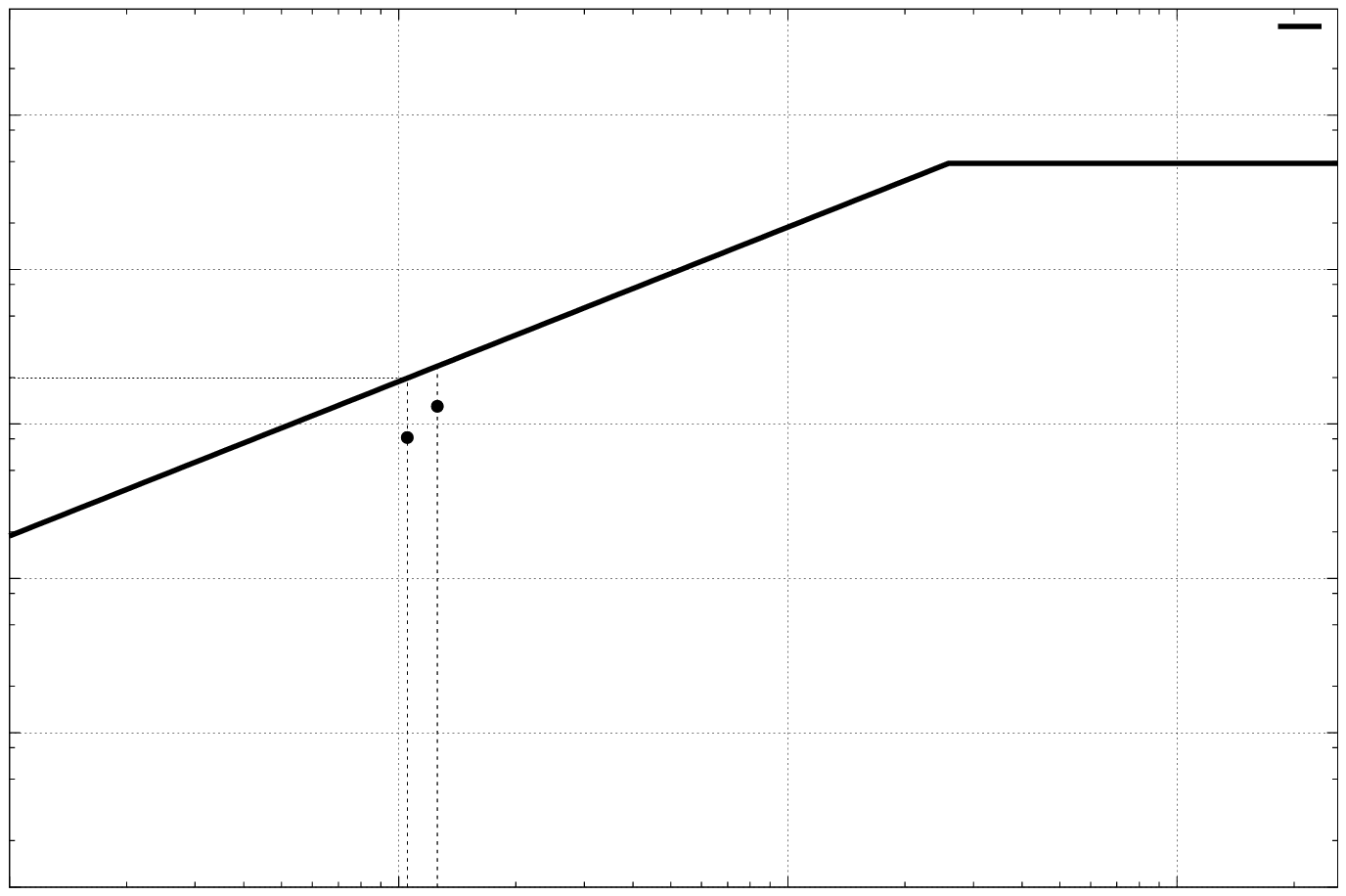}}%
    \gplfronttext
  \end{picture}%
\endgroup

\subsection*{GELU activation} \label{gelu_plots}

\begingroup
  \fontfamily{Helvetica}%
  \selectfont
  \makeatletter
  \providecommand\color[2][]{%
    \GenericError{(gnuplot) \space\space\space\@spaces}{%
      Package color not loaded in conjunction with
      terminal option `colourtext'%
    }{See the gnuplot documentation for explanation.%
    }{Either use 'blacktext' in gnuplot or load the package
      color.sty in LaTeX.}%
    \renewcommand\color[2][]{}%
  }%
  \providecommand\includegraphics[2][]{%
    \GenericError{(gnuplot) \space\space\space\@spaces}{%
      Package graphicx or graphics not loaded%
    }{See the gnuplot documentation for explanation.%
    }{The gnuplot epslatex terminal needs graphicx.sty or graphics.sty.}%
    \renewcommand\includegraphics[2][]{}%
  }%
  \providecommand\rotatebox[2]{#2}%
  \@ifundefined{ifGPcolor}{%
    \newif\ifGPcolor
    \GPcolorfalse
  }{}%
  \@ifundefined{ifGPblacktext}{%
    \newif\ifGPblacktext
    \GPblacktexttrue
  }{}%
  \let\gplgaddtomacro\g@addto@macro
  \gdef\gplbacktext{}%
  \gdef\gplfronttext{}%
  \makeatother
  \ifGPblacktext
    \def\colorrgb#1{}%
    \def\colorgray#1{}%
  \else
    \ifGPcolor
      \def\colorrgb#1{\color[rgb]{#1}}%
      \def\colorgray#1{\color[gray]{#1}}%
      \expandafter\def\csname LTw\endcsname{\color{white}}%
      \expandafter\def\csname LTb\endcsname{\color{black}}%
      \expandafter\def\csname LTa\endcsname{\color{black}}%
      \expandafter\def\csname LT0\endcsname{\color[rgb]{1,0,0}}%
      \expandafter\def\csname LT1\endcsname{\color[rgb]{0,1,0}}%
      \expandafter\def\csname LT2\endcsname{\color[rgb]{0,0,1}}%
      \expandafter\def\csname LT3\endcsname{\color[rgb]{1,0,1}}%
      \expandafter\def\csname LT4\endcsname{\color[rgb]{0,1,1}}%
      \expandafter\def\csname LT5\endcsname{\color[rgb]{1,1,0}}%
      \expandafter\def\csname LT6\endcsname{\color[rgb]{0,0,0}}%
      \expandafter\def\csname LT7\endcsname{\color[rgb]{1,0.3,0}}%
      \expandafter\def\csname LT8\endcsname{\color[rgb]{0.5,0.5,0.5}}%
    \else
      \def\colorrgb#1{\color{black}}%
      \def\colorgray#1{\color[gray]{#1}}%
      \expandafter\def\csname LTw\endcsname{\color{white}}%
      \expandafter\def\csname LTb\endcsname{\color{black}}%
      \expandafter\def\csname LTa\endcsname{\color{black}}%
      \expandafter\def\csname LT0\endcsname{\color{black}}%
      \expandafter\def\csname LT1\endcsname{\color{black}}%
      \expandafter\def\csname LT2\endcsname{\color{black}}%
      \expandafter\def\csname LT3\endcsname{\color{black}}%
      \expandafter\def\csname LT4\endcsname{\color{black}}%
      \expandafter\def\csname LT5\endcsname{\color{black}}%
      \expandafter\def\csname LT6\endcsname{\color{black}}%
      \expandafter\def\csname LT7\endcsname{\color{black}}%
      \expandafter\def\csname LT8\endcsname{\color{black}}%
    \fi
  \fi
    \setlength{\unitlength}{0.0500bp}%
    \ifx\gptboxheight\undefined%
      \newlength{\gptboxheight}%
      \newlength{\gptboxwidth}%
      \newsavebox{\gptboxtext}%
    \fi%
    \setlength{\fboxrule}{0.5pt}%
    \setlength{\fboxsep}{1pt}%
\begin{picture}(8502.00,5668.00)%
    \gplgaddtomacro\gplbacktext{%
      \csname LTb\endcsname%
      \put(440,256){\makebox(0,0)[r]{\strut{}\tiny $0.1$}}%
      \csname LTb\endcsname%
      \put(440,1515){\makebox(0,0)[r]{\strut{}\tiny $1$}}%
      \csname LTb\endcsname%
      \put(440,2774){\makebox(0,0)[r]{\strut{}\tiny $10$}}%
      \csname LTb\endcsname%
      \put(440,4033){\makebox(0,0)[r]{\strut{}\tiny $100$}}%
      \csname LTb\endcsname%
      \put(440,5292){\makebox(0,0)[r]{\strut{}\tiny $1000$}}%
      \csname LTb\endcsname%
      \put(488,176){\makebox(0,0){\strut{}\tiny $0.1$}}%
      \csname LTb\endcsname%
      \put(3013,176){\makebox(0,0){\strut{}\tiny $1$}}%
      \csname LTb\endcsname%
      \put(5537,176){\makebox(0,0){\strut{}\tiny $10$}}%
      \csname LTb\endcsname%
      \put(8062,176){\makebox(0,0){\strut{}\tiny $100$}}%
      \colorrgb{0.00,0.00,0.00}%
      \put(5832,4268){\makebox(0,0)[l]{\strut{}\tiny compute bound (Peak Runtime Compute: 100\%)}}%
      \put(4160,453){\rotatebox{90}{\makebox(0,0)[l]{\strut{}\tiny  NCHW GELU 256x96x55x55 (cold caches)}}}%
      \put(3425,2962){\makebox(0,0)[l]{\strut{}\tiny RC: 16.22\%}}%
      \csname LTb\endcsname%
      \put(850,3456){\makebox(0,0)[l]{\strut{}\tiny Attainable RC: 24.72 \%}}%
      \colorrgb{0.00,0.00,0.00}%
      \put(4474,453){\rotatebox{90}{\makebox(0,0)[l]{\strut{}\tiny  NCHW GELU 256x96x55x55 (warm caches)}}}%
      \put(4503,3255){\makebox(0,0)[l]{\strut{}\tiny RC: 17.93\%}}%
      \put(4245,453){\rotatebox{90}{\makebox(0,0)[l]{\strut{}\tiny  NCHW16C GELU 256x96x55x55 (cold caches)}}}%
      \put(3425,2885){\makebox(0,0)[l]{\strut{}\tiny RC: 14.08\%}}%
      \put(4401,453){\rotatebox{90}{\makebox(0,0)[l]{\strut{}\tiny  NCHW16C GELU 256x96x55x55 (warm caches)}}}%
      \put(4501,3166){\makebox(0,0)[l]{\strut{}\tiny RC: 16.83\%}}%
    }%
    \gplgaddtomacro\gplfronttext{%
      \csname LTb\endcsname%
      \put(64,2841){\rotatebox{-270}{\makebox(0,0){\strut{}\tiny Atteinable GFLOPS/s}}}%
      \put(4422,56){\makebox(0,0){\strut{}\tiny Arithmetic Intensity [FLOPS/Byte]}}%
      \put(4422,5547){\makebox(0,0){\strut{}\tiny Intel(R) Xeon(R) Gold 6248 CPU @ 2.50GHz; (single core)}}%
      \csname LTb\endcsname%
      \put(7958,5324){\makebox(0,0)[r]{\strut{}\small Roofline}}%
    }%
    \gplbacktext
    \put(0,0){\includegraphics{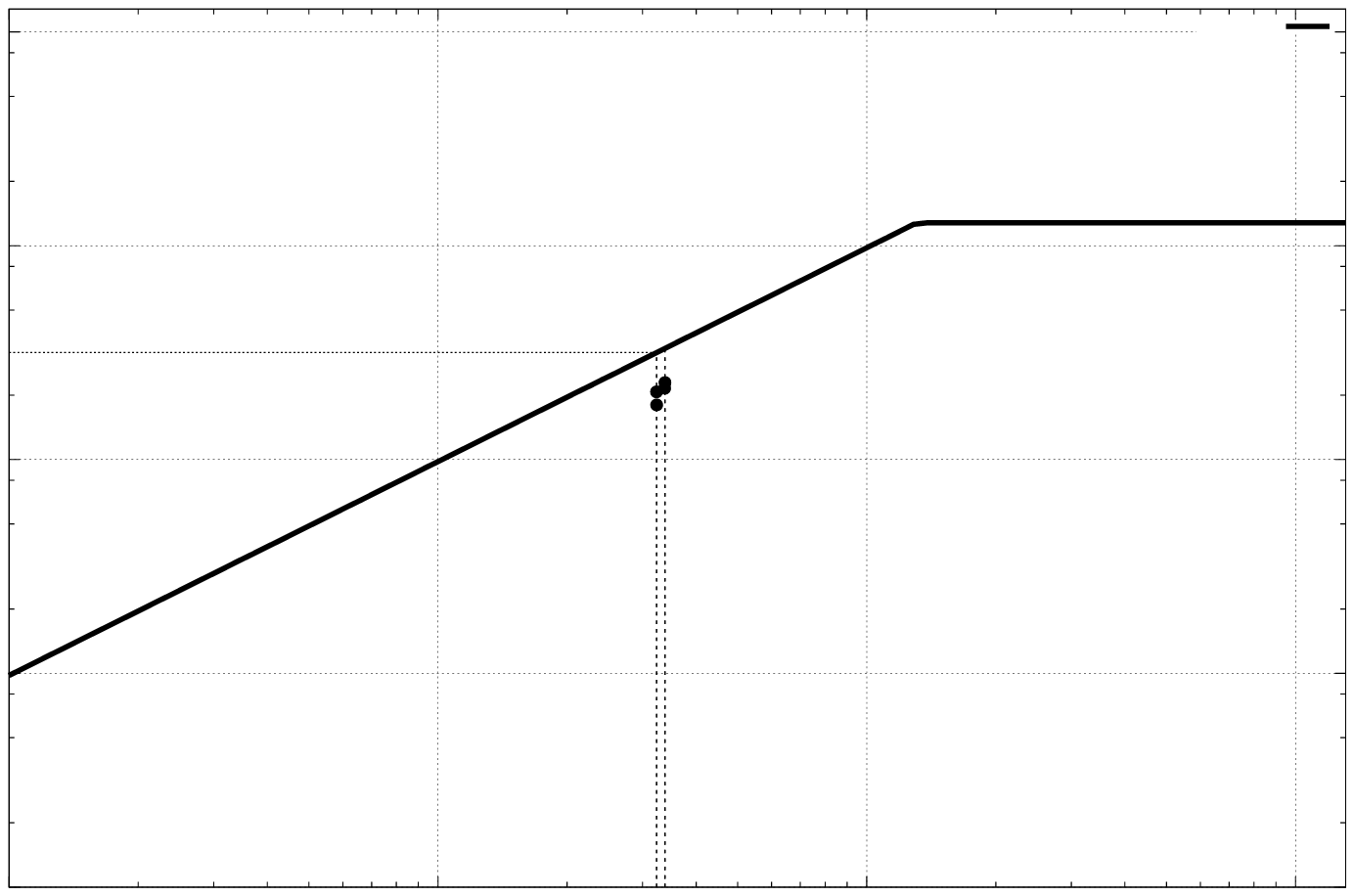}}%
    \gplfronttext
  \end{picture}%
\endgroup

\begingroup
  \fontfamily{Helvetica}%
  \selectfont
  \makeatletter
  \providecommand\color[2][]{%
    \GenericError{(gnuplot) \space\space\space\@spaces}{%
      Package color not loaded in conjunction with
      terminal option `colourtext'%
    }{See the gnuplot documentation for explanation.%
    }{Either use 'blacktext' in gnuplot or load the package
      color.sty in LaTeX.}%
    \renewcommand\color[2][]{}%
  }%
  \providecommand\includegraphics[2][]{%
    \GenericError{(gnuplot) \space\space\space\@spaces}{%
      Package graphicx or graphics not loaded%
    }{See the gnuplot documentation for explanation.%
    }{The gnuplot epslatex terminal needs graphicx.sty or graphics.sty.}%
    \renewcommand\includegraphics[2][]{}%
  }%
  \providecommand\rotatebox[2]{#2}%
  \@ifundefined{ifGPcolor}{%
    \newif\ifGPcolor
    \GPcolorfalse
  }{}%
  \@ifundefined{ifGPblacktext}{%
    \newif\ifGPblacktext
    \GPblacktexttrue
  }{}%
  \let\gplgaddtomacro\g@addto@macro
  \gdef\gplbacktext{}%
  \gdef\gplfronttext{}%
  \makeatother
  \ifGPblacktext
    \def\colorrgb#1{}%
    \def\colorgray#1{}%
  \else
    \ifGPcolor
      \def\colorrgb#1{\color[rgb]{#1}}%
      \def\colorgray#1{\color[gray]{#1}}%
      \expandafter\def\csname LTw\endcsname{\color{white}}%
      \expandafter\def\csname LTb\endcsname{\color{black}}%
      \expandafter\def\csname LTa\endcsname{\color{black}}%
      \expandafter\def\csname LT0\endcsname{\color[rgb]{1,0,0}}%
      \expandafter\def\csname LT1\endcsname{\color[rgb]{0,1,0}}%
      \expandafter\def\csname LT2\endcsname{\color[rgb]{0,0,1}}%
      \expandafter\def\csname LT3\endcsname{\color[rgb]{1,0,1}}%
      \expandafter\def\csname LT4\endcsname{\color[rgb]{0,1,1}}%
      \expandafter\def\csname LT5\endcsname{\color[rgb]{1,1,0}}%
      \expandafter\def\csname LT6\endcsname{\color[rgb]{0,0,0}}%
      \expandafter\def\csname LT7\endcsname{\color[rgb]{1,0.3,0}}%
      \expandafter\def\csname LT8\endcsname{\color[rgb]{0.5,0.5,0.5}}%
    \else
      \def\colorrgb#1{\color{black}}%
      \def\colorgray#1{\color[gray]{#1}}%
      \expandafter\def\csname LTw\endcsname{\color{white}}%
      \expandafter\def\csname LTb\endcsname{\color{black}}%
      \expandafter\def\csname LTa\endcsname{\color{black}}%
      \expandafter\def\csname LT0\endcsname{\color{black}}%
      \expandafter\def\csname LT1\endcsname{\color{black}}%
      \expandafter\def\csname LT2\endcsname{\color{black}}%
      \expandafter\def\csname LT3\endcsname{\color{black}}%
      \expandafter\def\csname LT4\endcsname{\color{black}}%
      \expandafter\def\csname LT5\endcsname{\color{black}}%
      \expandafter\def\csname LT6\endcsname{\color{black}}%
      \expandafter\def\csname LT7\endcsname{\color{black}}%
      \expandafter\def\csname LT8\endcsname{\color{black}}%
    \fi
  \fi
    \setlength{\unitlength}{0.0500bp}%
    \ifx\gptboxheight\undefined%
      \newlength{\gptboxheight}%
      \newlength{\gptboxwidth}%
      \newsavebox{\gptboxtext}%
    \fi%
    \setlength{\fboxrule}{0.5pt}%
    \setlength{\fboxsep}{1pt}%
\begin{picture}(8502.00,5668.00)%
    \gplgaddtomacro\gplbacktext{%
      \csname LTb\endcsname%
      \put(488,256){\makebox(0,0)[r]{\strut{}\tiny $0.1$}}%
      \csname LTb\endcsname%
      \put(488,1212){\makebox(0,0)[r]{\strut{}\tiny $1$}}%
      \csname LTb\endcsname%
      \put(488,2168){\makebox(0,0)[r]{\strut{}\tiny $10$}}%
      \csname LTb\endcsname%
      \put(488,3125){\makebox(0,0)[r]{\strut{}\tiny $100$}}%
      \csname LTb\endcsname%
      \put(488,4081){\makebox(0,0)[r]{\strut{}\tiny $1000$}}%
      \csname LTb\endcsname%
      \put(488,5037){\makebox(0,0)[r]{\strut{}\tiny $10000$}}%
      \csname LTb\endcsname%
      \put(536,176){\makebox(0,0){\strut{}\tiny $0.1$}}%
      \csname LTb\endcsname%
      \put(2810,176){\makebox(0,0){\strut{}\tiny $1$}}%
      \csname LTb\endcsname%
      \put(5085,176){\makebox(0,0){\strut{}\tiny $10$}}%
      \csname LTb\endcsname%
      \put(7359,176){\makebox(0,0){\strut{}\tiny $100$}}%
      \colorrgb{0.00,0.00,0.00}%
      \put(6083,4546){\makebox(0,0)[l]{\strut{}\tiny compute bound (Peak Runtime Compute: 100\%)}}%
      \put(4247,453){\rotatebox{90}{\makebox(0,0)[l]{\strut{}\tiny  NCHW GELU 256x96x55x55 (cold caches)}}}%
      \put(3689,3373){\makebox(0,0)[l]{\strut{}\tiny RC: 10.47\%}}%
      \csname LTb\endcsname%
      \put(850,3792){\makebox(0,0)[l]{\strut{}\tiny Attainable RC: 17.72 \%}}%
      \colorrgb{0.00,0.00,0.00}%
      \put(4486,453){\rotatebox{90}{\makebox(0,0)[l]{\strut{}\tiny  NCHW GELU 256x96x55x55 (warm caches)}}}%
      \put(4557,3476){\makebox(0,0)[l]{\strut{}\tiny RC: 10.13\%}}%
      \put(4323,453){\rotatebox{90}{\makebox(0,0)[l]{\strut{}\tiny  NCHW16C GELU 256x96x55x55 (cold caches)}}}%
      \put(3689,3446){\makebox(0,0)[l]{\strut{}\tiny RC: 10.59\%}}%
      \put(4564,453){\rotatebox{90}{\makebox(0,0)[l]{\strut{}\tiny  NCHW16C GELU 256x96x55x55 (warm caches)}}}%
      \put(4564,3545){\makebox(0,0)[l]{\strut{}\tiny RC: 10.14\%}}%
    }%
    \gplgaddtomacro\gplfronttext{%
      \csname LTb\endcsname%
      \put(64,2841){\rotatebox{-270}{\makebox(0,0){\strut{}\tiny Atteinable GFLOPS/s}}}%
      \put(4446,56){\makebox(0,0){\strut{}\tiny Arithmetic Intensity [FLOPS/Byte]}}%
      \put(4446,5547){\makebox(0,0){\strut{}\tiny Intel(R) Xeon(R) Gold 6248 CPU @ 2.50GHz; (single socket)}}%
      \csname LTb\endcsname%
      \put(7958,5324){\makebox(0,0)[r]{\strut{}\small Roofline}}%
    }%
    \gplbacktext
    \put(0,0){\includegraphics{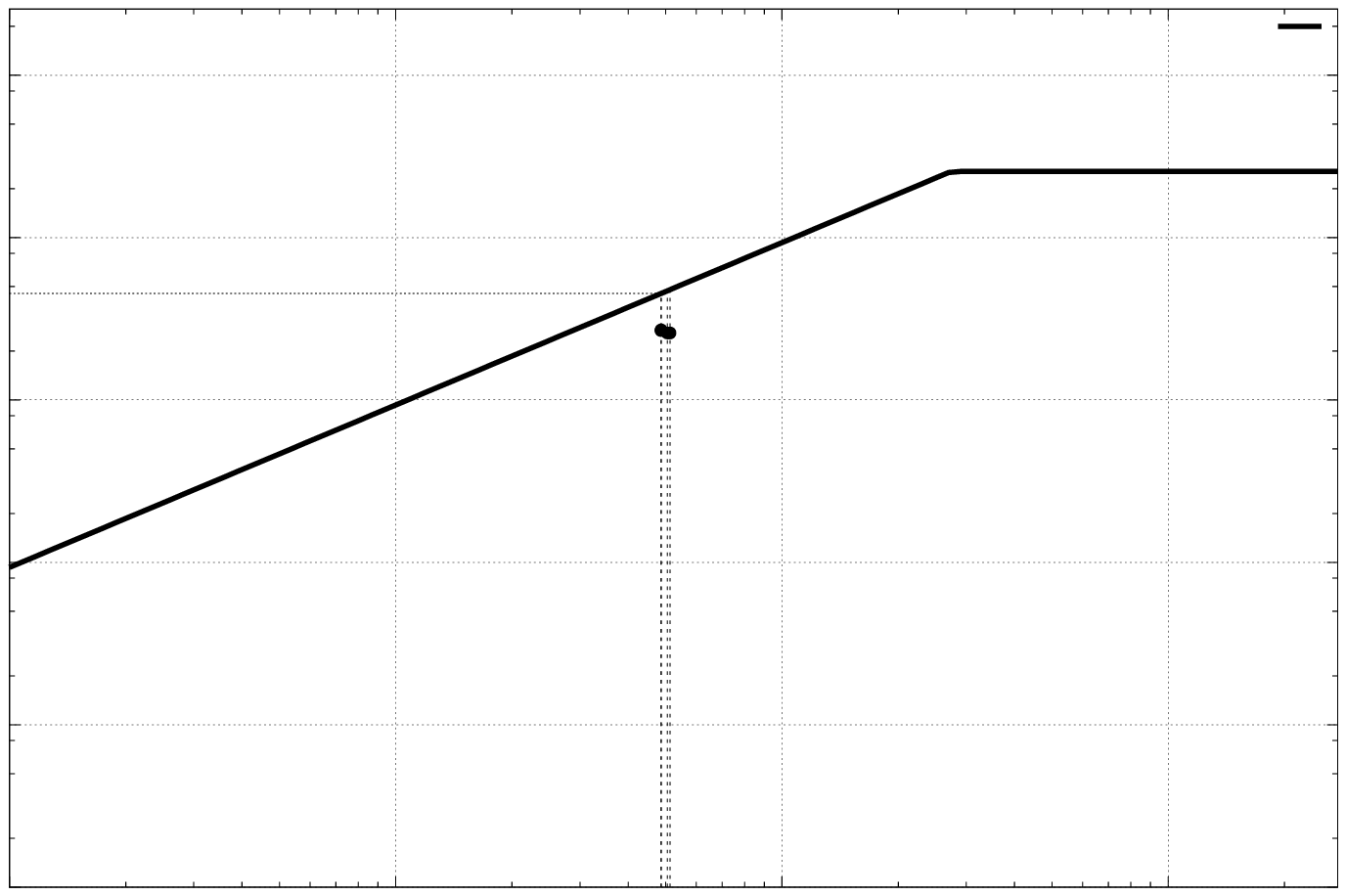}}%
    \gplfronttext
  \end{picture}%
\endgroup

\begingroup
  \fontfamily{Helvetica}%
  \selectfont
  \makeatletter
  \providecommand\color[2][]{%
    \GenericError{(gnuplot) \space\space\space\@spaces}{%
      Package color not loaded in conjunction with
      terminal option `colourtext'%
    }{See the gnuplot documentation for explanation.%
    }{Either use 'blacktext' in gnuplot or load the package
      color.sty in LaTeX.}%
    \renewcommand\color[2][]{}%
  }%
  \providecommand\includegraphics[2][]{%
    \GenericError{(gnuplot) \space\space\space\@spaces}{%
      Package graphicx or graphics not loaded%
    }{See the gnuplot documentation for explanation.%
    }{The gnuplot epslatex terminal needs graphicx.sty or graphics.sty.}%
    \renewcommand\includegraphics[2][]{}%
  }%
  \providecommand\rotatebox[2]{#2}%
  \@ifundefined{ifGPcolor}{%
    \newif\ifGPcolor
    \GPcolorfalse
  }{}%
  \@ifundefined{ifGPblacktext}{%
    \newif\ifGPblacktext
    \GPblacktexttrue
  }{}%
  \let\gplgaddtomacro\g@addto@macro
  \gdef\gplbacktext{}%
  \gdef\gplfronttext{}%
  \makeatother
  \ifGPblacktext
    \def\colorrgb#1{}%
    \def\colorgray#1{}%
  \else
    \ifGPcolor
      \def\colorrgb#1{\color[rgb]{#1}}%
      \def\colorgray#1{\color[gray]{#1}}%
      \expandafter\def\csname LTw\endcsname{\color{white}}%
      \expandafter\def\csname LTb\endcsname{\color{black}}%
      \expandafter\def\csname LTa\endcsname{\color{black}}%
      \expandafter\def\csname LT0\endcsname{\color[rgb]{1,0,0}}%
      \expandafter\def\csname LT1\endcsname{\color[rgb]{0,1,0}}%
      \expandafter\def\csname LT2\endcsname{\color[rgb]{0,0,1}}%
      \expandafter\def\csname LT3\endcsname{\color[rgb]{1,0,1}}%
      \expandafter\def\csname LT4\endcsname{\color[rgb]{0,1,1}}%
      \expandafter\def\csname LT5\endcsname{\color[rgb]{1,1,0}}%
      \expandafter\def\csname LT6\endcsname{\color[rgb]{0,0,0}}%
      \expandafter\def\csname LT7\endcsname{\color[rgb]{1,0.3,0}}%
      \expandafter\def\csname LT8\endcsname{\color[rgb]{0.5,0.5,0.5}}%
    \else
      \def\colorrgb#1{\color{black}}%
      \def\colorgray#1{\color[gray]{#1}}%
      \expandafter\def\csname LTw\endcsname{\color{white}}%
      \expandafter\def\csname LTb\endcsname{\color{black}}%
      \expandafter\def\csname LTa\endcsname{\color{black}}%
      \expandafter\def\csname LT0\endcsname{\color{black}}%
      \expandafter\def\csname LT1\endcsname{\color{black}}%
      \expandafter\def\csname LT2\endcsname{\color{black}}%
      \expandafter\def\csname LT3\endcsname{\color{black}}%
      \expandafter\def\csname LT4\endcsname{\color{black}}%
      \expandafter\def\csname LT5\endcsname{\color{black}}%
      \expandafter\def\csname LT6\endcsname{\color{black}}%
      \expandafter\def\csname LT7\endcsname{\color{black}}%
      \expandafter\def\csname LT8\endcsname{\color{black}}%
    \fi
  \fi
    \setlength{\unitlength}{0.0500bp}%
    \ifx\gptboxheight\undefined%
      \newlength{\gptboxheight}%
      \newlength{\gptboxwidth}%
      \newsavebox{\gptboxtext}%
    \fi%
    \setlength{\fboxrule}{0.5pt}%
    \setlength{\fboxsep}{1pt}%
\begin{picture}(8502.00,5668.00)%
    \gplgaddtomacro\gplbacktext{%
      \csname LTb\endcsname%
      \put(488,256){\makebox(0,0)[r]{\strut{}\tiny $0.1$}}%
      \csname LTb\endcsname%
      \put(488,1168){\makebox(0,0)[r]{\strut{}\tiny $1$}}%
      \csname LTb\endcsname%
      \put(488,2079){\makebox(0,0)[r]{\strut{}\tiny $10$}}%
      \csname LTb\endcsname%
      \put(488,2991){\makebox(0,0)[r]{\strut{}\tiny $100$}}%
      \csname LTb\endcsname%
      \put(488,3902){\makebox(0,0)[r]{\strut{}\tiny $1000$}}%
      \csname LTb\endcsname%
      \put(488,4814){\makebox(0,0)[r]{\strut{}\tiny $10000$}}%
      \csname LTb\endcsname%
      \put(536,176){\makebox(0,0){\strut{}\tiny $0.1$}}%
      \csname LTb\endcsname%
      \put(2842,176){\makebox(0,0){\strut{}\tiny $1$}}%
      \csname LTb\endcsname%
      \put(5148,176){\makebox(0,0){\strut{}\tiny $10$}}%
      \csname LTb\endcsname%
      \put(7454,176){\makebox(0,0){\strut{}\tiny $100$}}%
      \colorrgb{0.00,0.00,0.00}%
      \put(6051,4588){\makebox(0,0)[l]{\strut{}\tiny compute bound (Peak Runtime Compute: 100\%)}}%
      \put(4285,453){\rotatebox{90}{\makebox(0,0)[l]{\strut{}\tiny  NCHW GELU 256x96x55x55 (cold caches)}}}%
      \put(3718,3437){\makebox(0,0)[l]{\strut{}\tiny RC: 7.47\%}}%
      \csname LTb\endcsname%
      \put(850,3905){\makebox(0,0)[l]{\strut{}\tiny Attainable RC: 19.47 \%}}%
      \colorrgb{0.00,0.00,0.00}%
      \put(4572,453){\rotatebox{90}{\makebox(0,0)[l]{\strut{}\tiny  NCHW GELU 256x96x55x55 (warm caches)}}}%
      \put(4507,3542){\makebox(0,0)[l]{\strut{}\tiny RC: 9.51\%}}%
      \put(4364,453){\rotatebox{90}{\makebox(0,0)[l]{\strut{}\tiny  NCHW16C GELU 256x96x55x55 (cold caches)}}}%
      \put(3722,3507){\makebox(0,0)[l]{\strut{}\tiny RC: 7.83\%}}%
      \put(4500,453){\rotatebox{90}{\makebox(0,0)[l]{\strut{}\tiny  NCHW16C GELU 256x96x55x55 (warm caches)}}}%
      \put(4500,3618){\makebox(0,0)[l]{\strut{}\tiny RC: 10.36\%}}%
    }%
    \gplgaddtomacro\gplfronttext{%
      \csname LTb\endcsname%
      \put(64,2841){\rotatebox{-270}{\makebox(0,0){\strut{}\tiny Atteinable GFLOPS/s}}}%
      \put(4446,56){\makebox(0,0){\strut{}\tiny Arithmetic Intensity [FLOPS/Byte]}}%
      \put(4446,5547){\makebox(0,0){\strut{}\tiny Intel(R) Xeon(R) Gold 6248 CPU @ 2.50GHz; (two sockets)}}%
      \csname LTb\endcsname%
      \put(7958,5324){\makebox(0,0)[r]{\strut{}\small Roofline}}%
    }%
    \gplbacktext
    \put(0,0){\includegraphics{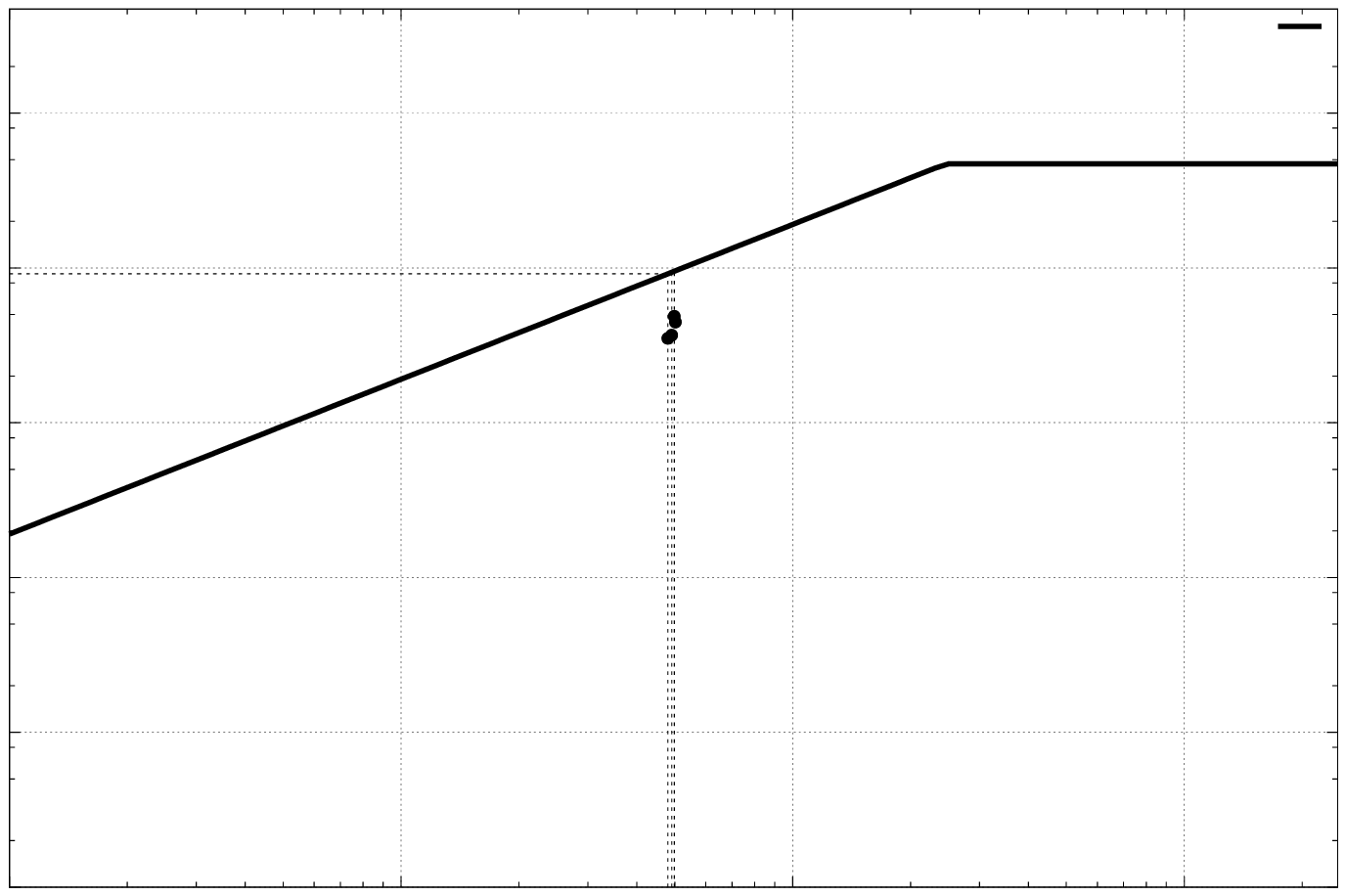}}%
    \gplfronttext
  \end{picture}%
\endgroup

\subsection*{Inner Product (Fully Connected) } \label{fc_plots}

\begingroup
  \fontfamily{Helvetica}%
  \selectfont
  \makeatletter
  \providecommand\color[2][]{%
    \GenericError{(gnuplot) \space\space\space\@spaces}{%
      Package color not loaded in conjunction with
      terminal option `colourtext'%
    }{See the gnuplot documentation for explanation.%
    }{Either use 'blacktext' in gnuplot or load the package
      color.sty in LaTeX.}%
    \renewcommand\color[2][]{}%
  }%
  \providecommand\includegraphics[2][]{%
    \GenericError{(gnuplot) \space\space\space\@spaces}{%
      Package graphicx or graphics not loaded%
    }{See the gnuplot documentation for explanation.%
    }{The gnuplot epslatex terminal needs graphicx.sty or graphics.sty.}%
    \renewcommand\includegraphics[2][]{}%
  }%
  \providecommand\rotatebox[2]{#2}%
  \@ifundefined{ifGPcolor}{%
    \newif\ifGPcolor
    \GPcolorfalse
  }{}%
  \@ifundefined{ifGPblacktext}{%
    \newif\ifGPblacktext
    \GPblacktexttrue
  }{}%
  \let\gplgaddtomacro\g@addto@macro
  \gdef\gplbacktext{}%
  \gdef\gplfronttext{}%
  \makeatother
  \ifGPblacktext
    \def\colorrgb#1{}%
    \def\colorgray#1{}%
  \else
    \ifGPcolor
      \def\colorrgb#1{\color[rgb]{#1}}%
      \def\colorgray#1{\color[gray]{#1}}%
      \expandafter\def\csname LTw\endcsname{\color{white}}%
      \expandafter\def\csname LTb\endcsname{\color{black}}%
      \expandafter\def\csname LTa\endcsname{\color{black}}%
      \expandafter\def\csname LT0\endcsname{\color[rgb]{1,0,0}}%
      \expandafter\def\csname LT1\endcsname{\color[rgb]{0,1,0}}%
      \expandafter\def\csname LT2\endcsname{\color[rgb]{0,0,1}}%
      \expandafter\def\csname LT3\endcsname{\color[rgb]{1,0,1}}%
      \expandafter\def\csname LT4\endcsname{\color[rgb]{0,1,1}}%
      \expandafter\def\csname LT5\endcsname{\color[rgb]{1,1,0}}%
      \expandafter\def\csname LT6\endcsname{\color[rgb]{0,0,0}}%
      \expandafter\def\csname LT7\endcsname{\color[rgb]{1,0.3,0}}%
      \expandafter\def\csname LT8\endcsname{\color[rgb]{0.5,0.5,0.5}}%
    \else
      \def\colorrgb#1{\color{black}}%
      \def\colorgray#1{\color[gray]{#1}}%
      \expandafter\def\csname LTw\endcsname{\color{white}}%
      \expandafter\def\csname LTb\endcsname{\color{black}}%
      \expandafter\def\csname LTa\endcsname{\color{black}}%
      \expandafter\def\csname LT0\endcsname{\color{black}}%
      \expandafter\def\csname LT1\endcsname{\color{black}}%
      \expandafter\def\csname LT2\endcsname{\color{black}}%
      \expandafter\def\csname LT3\endcsname{\color{black}}%
      \expandafter\def\csname LT4\endcsname{\color{black}}%
      \expandafter\def\csname LT5\endcsname{\color{black}}%
      \expandafter\def\csname LT6\endcsname{\color{black}}%
      \expandafter\def\csname LT7\endcsname{\color{black}}%
      \expandafter\def\csname LT8\endcsname{\color{black}}%
    \fi
  \fi
    \setlength{\unitlength}{0.0500bp}%
    \ifx\gptboxheight\undefined%
      \newlength{\gptboxheight}%
      \newlength{\gptboxwidth}%
      \newsavebox{\gptboxtext}%
    \fi%
    \setlength{\fboxrule}{0.5pt}%
    \setlength{\fboxsep}{1pt}%
\begin{picture}(8502.00,5668.00)%
    \gplgaddtomacro\gplbacktext{%
      \csname LTb\endcsname%
      \put(488,256){\makebox(0,0)[r]{\strut{}\tiny $0.1$}}%
      \csname LTb\endcsname%
      \put(488,1212){\makebox(0,0)[r]{\strut{}\tiny $1$}}%
      \csname LTb\endcsname%
      \put(488,2168){\makebox(0,0)[r]{\strut{}\tiny $10$}}%
      \csname LTb\endcsname%
      \put(488,3125){\makebox(0,0)[r]{\strut{}\tiny $100$}}%
      \csname LTb\endcsname%
      \put(488,4081){\makebox(0,0)[r]{\strut{}\tiny $1000$}}%
      \csname LTb\endcsname%
      \put(488,5037){\makebox(0,0)[r]{\strut{}\tiny $10000$}}%
      \csname LTb\endcsname%
      \put(536,176){\makebox(0,0){\strut{}\tiny $0.1$}}%
      \csname LTb\endcsname%
      \put(2546,176){\makebox(0,0){\strut{}\tiny $1$}}%
      \csname LTb\endcsname%
      \put(4556,176){\makebox(0,0){\strut{}\tiny $10$}}%
      \csname LTb\endcsname%
      \put(6567,176){\makebox(0,0){\strut{}\tiny $100$}}%
      \colorrgb{0.00,0.00,0.00}%
      \put(5419,4546){\makebox(0,0)[l]{\strut{}\tiny compute bound (Peak Runtime Compute: 100\%)}}%
      \put(6302,453){\rotatebox{90}{\makebox(0,0)[l]{\strut{}\tiny  NCHW FC 256x4096x1x1 1000x1x1 (cold caches)}}}%
      \put(5776,4025){\makebox(0,0)[l]{\strut{}\tiny RC: 37.95\%}}%
      \csname LTb\endcsname%
      \put(850,4510){\makebox(0,0)[l]{\strut{}\tiny Attainable RC: 100.00 \%}}%
      \colorrgb{0.00,0.00,0.00}%
      \put(6618,453){\rotatebox{90}{\makebox(0,0)[l]{\strut{}\tiny  NCHW FC 256x4096x1x1 1000x1x1 (warm caches)}}}%
      \put(6680,4061){\makebox(0,0)[l]{\strut{}\tiny RC: 41.38\%}}%
      \put(850,397){\makebox(0,0)[l]{\strut{}\tiny *Chosen data arrangement is suboptimal. Diffrent data arrangement produce better utilisation }}%
    }%
    \gplgaddtomacro\gplfronttext{%
      \csname LTb\endcsname%
      \put(64,2841){\rotatebox{-270}{\makebox(0,0){\strut{}\tiny Atteinable GFLOPS/s}}}%
      \put(4446,56){\makebox(0,0){\strut{}\tiny Arithmetic Intensity [FLOPS/Byte]}}%
      \put(4446,5547){\makebox(0,0){\strut{}\tiny Intel(R) Xeon(R) Gold 6248 CPU @ 2.50GHz; (single socket)}}%
      \csname LTb\endcsname%
      \put(7958,5324){\makebox(0,0)[r]{\strut{}\small Roofline}}%
    }%
    \gplbacktext
    \put(0,0){\includegraphics{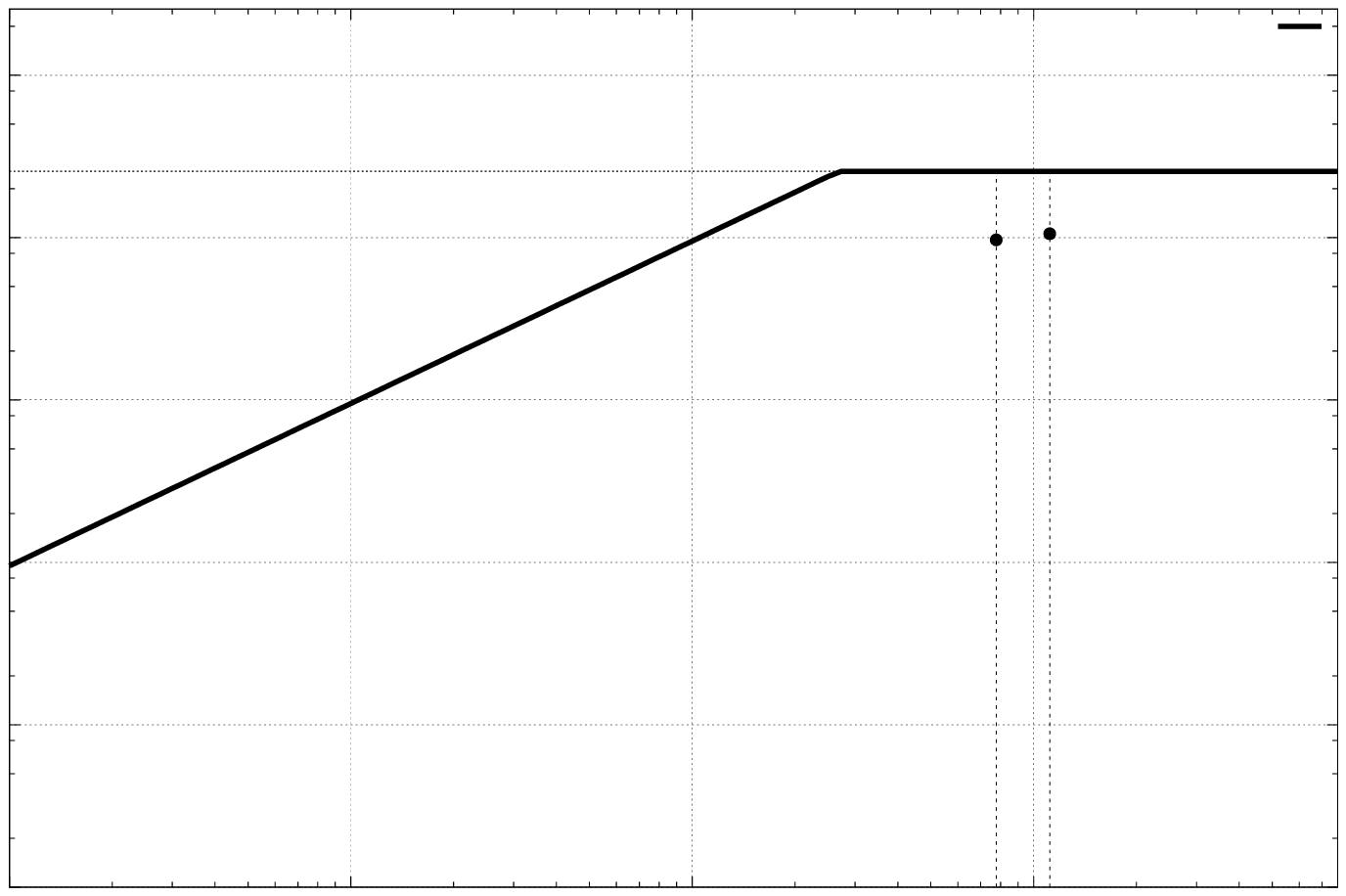}}%
    \gplfronttext
  \end{picture}%
\endgroup

\begingroup
  \fontfamily{Helvetica}%
  \selectfont
  \makeatletter
  \providecommand\color[2][]{%
    \GenericError{(gnuplot) \space\space\space\@spaces}{%
      Package color not loaded in conjunction with
      terminal option `colourtext'%
    }{See the gnuplot documentation for explanation.%
    }{Either use 'blacktext' in gnuplot or load the package
      color.sty in LaTeX.}%
    \renewcommand\color[2][]{}%
  }%
  \providecommand\includegraphics[2][]{%
    \GenericError{(gnuplot) \space\space\space\@spaces}{%
      Package graphicx or graphics not loaded%
    }{See the gnuplot documentation for explanation.%
    }{The gnuplot epslatex terminal needs graphicx.sty or graphics.sty.}%
    \renewcommand\includegraphics[2][]{}%
  }%
  \providecommand\rotatebox[2]{#2}%
  \@ifundefined{ifGPcolor}{%
    \newif\ifGPcolor
    \GPcolorfalse
  }{}%
  \@ifundefined{ifGPblacktext}{%
    \newif\ifGPblacktext
    \GPblacktexttrue
  }{}%
  \let\gplgaddtomacro\g@addto@macro
  \gdef\gplbacktext{}%
  \gdef\gplfronttext{}%
  \makeatother
  \ifGPblacktext
    \def\colorrgb#1{}%
    \def\colorgray#1{}%
  \else
    \ifGPcolor
      \def\colorrgb#1{\color[rgb]{#1}}%
      \def\colorgray#1{\color[gray]{#1}}%
      \expandafter\def\csname LTw\endcsname{\color{white}}%
      \expandafter\def\csname LTb\endcsname{\color{black}}%
      \expandafter\def\csname LTa\endcsname{\color{black}}%
      \expandafter\def\csname LT0\endcsname{\color[rgb]{1,0,0}}%
      \expandafter\def\csname LT1\endcsname{\color[rgb]{0,1,0}}%
      \expandafter\def\csname LT2\endcsname{\color[rgb]{0,0,1}}%
      \expandafter\def\csname LT3\endcsname{\color[rgb]{1,0,1}}%
      \expandafter\def\csname LT4\endcsname{\color[rgb]{0,1,1}}%
      \expandafter\def\csname LT5\endcsname{\color[rgb]{1,1,0}}%
      \expandafter\def\csname LT6\endcsname{\color[rgb]{0,0,0}}%
      \expandafter\def\csname LT7\endcsname{\color[rgb]{1,0.3,0}}%
      \expandafter\def\csname LT8\endcsname{\color[rgb]{0.5,0.5,0.5}}%
    \else
      \def\colorrgb#1{\color{black}}%
      \def\colorgray#1{\color[gray]{#1}}%
      \expandafter\def\csname LTw\endcsname{\color{white}}%
      \expandafter\def\csname LTb\endcsname{\color{black}}%
      \expandafter\def\csname LTa\endcsname{\color{black}}%
      \expandafter\def\csname LT0\endcsname{\color{black}}%
      \expandafter\def\csname LT1\endcsname{\color{black}}%
      \expandafter\def\csname LT2\endcsname{\color{black}}%
      \expandafter\def\csname LT3\endcsname{\color{black}}%
      \expandafter\def\csname LT4\endcsname{\color{black}}%
      \expandafter\def\csname LT5\endcsname{\color{black}}%
      \expandafter\def\csname LT6\endcsname{\color{black}}%
      \expandafter\def\csname LT7\endcsname{\color{black}}%
      \expandafter\def\csname LT8\endcsname{\color{black}}%
    \fi
  \fi
    \setlength{\unitlength}{0.0500bp}%
    \ifx\gptboxheight\undefined%
      \newlength{\gptboxheight}%
      \newlength{\gptboxwidth}%
      \newsavebox{\gptboxtext}%
    \fi%
    \setlength{\fboxrule}{0.5pt}%
    \setlength{\fboxsep}{1pt}%
\begin{picture}(8502.00,5668.00)%
    \gplgaddtomacro\gplbacktext{%
      \csname LTb\endcsname%
      \put(488,256){\makebox(0,0)[r]{\strut{}\tiny $0.1$}}%
      \csname LTb\endcsname%
      \put(488,1165){\makebox(0,0)[r]{\strut{}\tiny $1$}}%
      \csname LTb\endcsname%
      \put(488,2074){\makebox(0,0)[r]{\strut{}\tiny $10$}}%
      \csname LTb\endcsname%
      \put(488,2983){\makebox(0,0)[r]{\strut{}\tiny $100$}}%
      \csname LTb\endcsname%
      \put(488,3892){\makebox(0,0)[r]{\strut{}\tiny $1000$}}%
      \csname LTb\endcsname%
      \put(488,4801){\makebox(0,0)[r]{\strut{}\tiny $10000$}}%
      \csname LTb\endcsname%
      \put(536,176){\makebox(0,0){\strut{}\tiny $0.1$}}%
      \csname LTb\endcsname%
      \put(2590,176){\makebox(0,0){\strut{}\tiny $1$}}%
      \csname LTb\endcsname%
      \put(4644,176){\makebox(0,0){\strut{}\tiny $10$}}%
      \csname LTb\endcsname%
      \put(6697,176){\makebox(0,0){\strut{}\tiny $100$}}%
      \colorrgb{0.00,0.00,0.00}%
      \put(5488,4590){\makebox(0,0)[l]{\strut{}\tiny compute bound (Peak Runtime Compute: 100\%)}}%
      \put(5884,453){\rotatebox{90}{\makebox(0,0)[l]{\strut{}\tiny  NCHW FC 256x4096x1x1 1000x1x1 (cold caches)}}}%
      \put(5237,3840){\makebox(0,0)[l]{\strut{}\tiny RC: 19.95\%}}%
      \csname LTb\endcsname%
      \put(850,4556){\makebox(0,0)[l]{\strut{}\tiny Attainable RC: 100.00 \%}}%
      \colorrgb{0.00,0.00,0.00}%
      \put(6257,453){\rotatebox{90}{\makebox(0,0)[l]{\strut{}\tiny  NCHW FC 256x4096x1x1 1000x1x1 (warm caches)}}}%
      \put(6347,4045){\makebox(0,0)[l]{\strut{}\tiny RC: 33.52\%}}%
      \put(6801,357){\makebox(0,0)[l]{\strut{}\tiny RC - Runtime Compute}}%
    }%
    \gplgaddtomacro\gplfronttext{%
      \csname LTb\endcsname%
      \put(64,2841){\rotatebox{-270}{\makebox(0,0){\strut{}\tiny Atteinable GFLOPS/s}}}%
      \put(4446,56){\makebox(0,0){\strut{}\tiny Arithmetic Intensity [FLOPS/Byte]}}%
      \put(4446,5547){\makebox(0,0){\strut{}\tiny Intel(R) Xeon(R) Gold 6248 CPU @ 2.50GHz; (two sockets)}}%
      \csname LTb\endcsname%
      \put(7958,5324){\makebox(0,0)[r]{\strut{}\small Roofline}}%
    }%
    \gplbacktext
    \put(0,0){\includegraphics{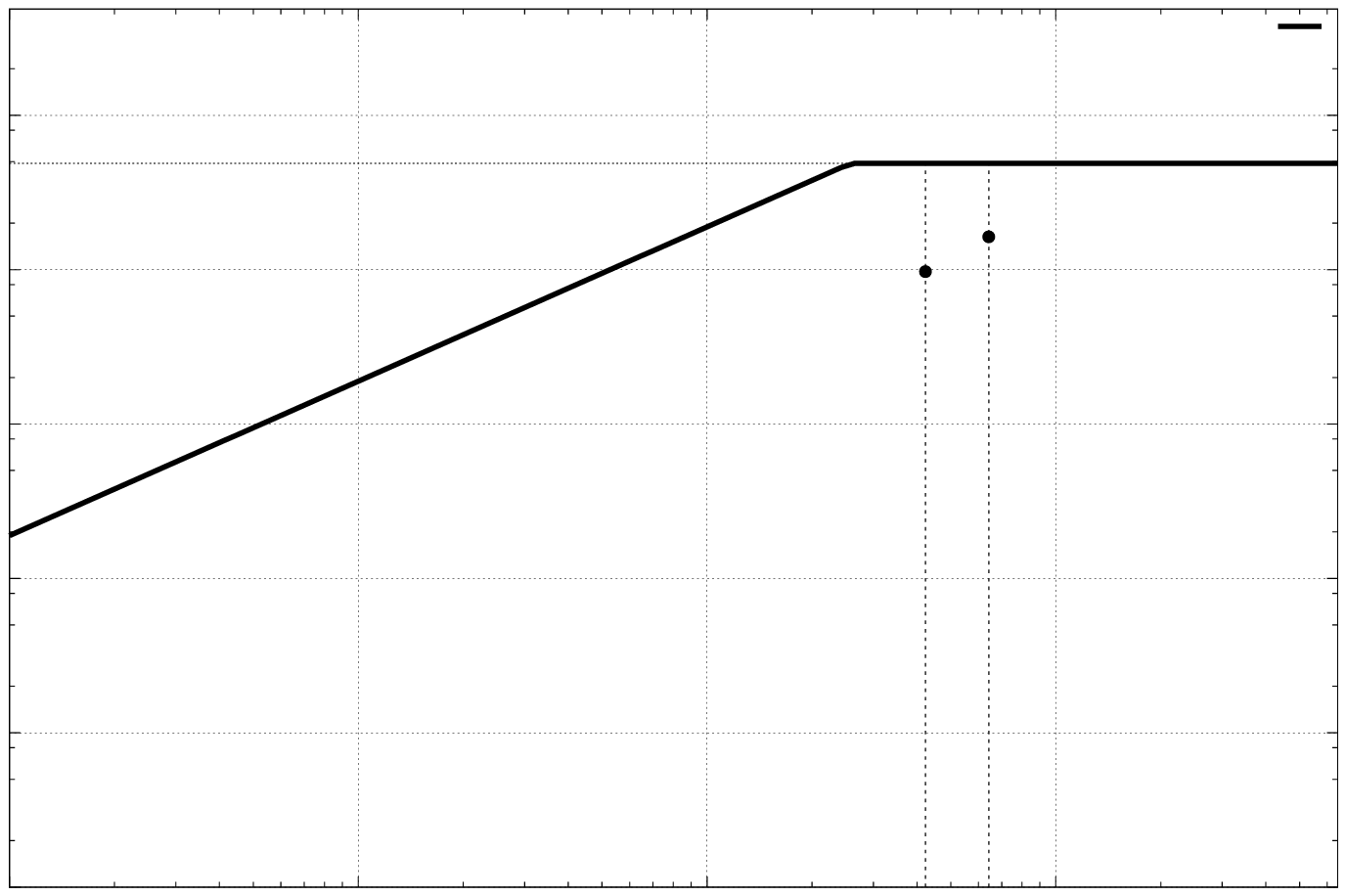}}%
    \gplfronttext
  \end{picture}%
\endgroup

\subsection*{Pooling (average) } \label{pool_plots}

\begingroup
  \fontfamily{Helvetica}%
  \selectfont
  \makeatletter
  \providecommand\color[2][]{%
    \GenericError{(gnuplot) \space\space\space\@spaces}{%
      Package color not loaded in conjunction with
      terminal option `colourtext'%
    }{See the gnuplot documentation for explanation.%
    }{Either use 'blacktext' in gnuplot or load the package
      color.sty in LaTeX.}%
    \renewcommand\color[2][]{}%
  }%
  \providecommand\includegraphics[2][]{%
    \GenericError{(gnuplot) \space\space\space\@spaces}{%
      Package graphicx or graphics not loaded%
    }{See the gnuplot documentation for explanation.%
    }{The gnuplot epslatex terminal needs graphicx.sty or graphics.sty.}%
    \renewcommand\includegraphics[2][]{}%
  }%
  \providecommand\rotatebox[2]{#2}%
  \@ifundefined{ifGPcolor}{%
    \newif\ifGPcolor
    \GPcolorfalse
  }{}%
  \@ifundefined{ifGPblacktext}{%
    \newif\ifGPblacktext
    \GPblacktexttrue
  }{}%
  \let\gplgaddtomacro\g@addto@macro
  \gdef\gplbacktext{}%
  \gdef\gplfronttext{}%
  \makeatother
  \ifGPblacktext
    \def\colorrgb#1{}%
    \def\colorgray#1{}%
  \else
    \ifGPcolor
      \def\colorrgb#1{\color[rgb]{#1}}%
      \def\colorgray#1{\color[gray]{#1}}%
      \expandafter\def\csname LTw\endcsname{\color{white}}%
      \expandafter\def\csname LTb\endcsname{\color{black}}%
      \expandafter\def\csname LTa\endcsname{\color{black}}%
      \expandafter\def\csname LT0\endcsname{\color[rgb]{1,0,0}}%
      \expandafter\def\csname LT1\endcsname{\color[rgb]{0,1,0}}%
      \expandafter\def\csname LT2\endcsname{\color[rgb]{0,0,1}}%
      \expandafter\def\csname LT3\endcsname{\color[rgb]{1,0,1}}%
      \expandafter\def\csname LT4\endcsname{\color[rgb]{0,1,1}}%
      \expandafter\def\csname LT5\endcsname{\color[rgb]{1,1,0}}%
      \expandafter\def\csname LT6\endcsname{\color[rgb]{0,0,0}}%
      \expandafter\def\csname LT7\endcsname{\color[rgb]{1,0.3,0}}%
      \expandafter\def\csname LT8\endcsname{\color[rgb]{0.5,0.5,0.5}}%
    \else
      \def\colorrgb#1{\color{black}}%
      \def\colorgray#1{\color[gray]{#1}}%
      \expandafter\def\csname LTw\endcsname{\color{white}}%
      \expandafter\def\csname LTb\endcsname{\color{black}}%
      \expandafter\def\csname LTa\endcsname{\color{black}}%
      \expandafter\def\csname LT0\endcsname{\color{black}}%
      \expandafter\def\csname LT1\endcsname{\color{black}}%
      \expandafter\def\csname LT2\endcsname{\color{black}}%
      \expandafter\def\csname LT3\endcsname{\color{black}}%
      \expandafter\def\csname LT4\endcsname{\color{black}}%
      \expandafter\def\csname LT5\endcsname{\color{black}}%
      \expandafter\def\csname LT6\endcsname{\color{black}}%
      \expandafter\def\csname LT7\endcsname{\color{black}}%
      \expandafter\def\csname LT8\endcsname{\color{black}}%
    \fi
  \fi
    \setlength{\unitlength}{0.0500bp}%
    \ifx\gptboxheight\undefined%
      \newlength{\gptboxheight}%
      \newlength{\gptboxwidth}%
      \newsavebox{\gptboxtext}%
    \fi%
    \setlength{\fboxrule}{0.5pt}%
    \setlength{\fboxsep}{1pt}%
\begin{picture}(8502.00,5668.00)%
    \gplgaddtomacro\gplbacktext{%
      \csname LTb\endcsname%
      \put(488,256){\makebox(0,0)[r]{\strut{}\tiny $0.1$}}%
      \csname LTb\endcsname%
      \put(488,1212){\makebox(0,0)[r]{\strut{}\tiny $1$}}%
      \csname LTb\endcsname%
      \put(488,2168){\makebox(0,0)[r]{\strut{}\tiny $10$}}%
      \csname LTb\endcsname%
      \put(488,3125){\makebox(0,0)[r]{\strut{}\tiny $100$}}%
      \csname LTb\endcsname%
      \put(488,4081){\makebox(0,0)[r]{\strut{}\tiny $1000$}}%
      \csname LTb\endcsname%
      \put(488,5037){\makebox(0,0)[r]{\strut{}\tiny $10000$}}%
      \csname LTb\endcsname%
      \put(536,176){\makebox(0,0){\strut{}\tiny $0.1$}}%
      \csname LTb\endcsname%
      \put(2816,176){\makebox(0,0){\strut{}\tiny $1$}}%
      \csname LTb\endcsname%
      \put(5096,176){\makebox(0,0){\strut{}\tiny $10$}}%
      \csname LTb\endcsname%
      \put(7376,176){\makebox(0,0){\strut{}\tiny $100$}}%
      \colorrgb{0.00,0.00,0.00}%
      \put(6077,4546){\makebox(0,0)[l]{\strut{}\tiny compute bound (Peak Runtime Compute: 100\%)}}%
      \put(3764,453){\rotatebox{90}{\makebox(0,0)[l]{\strut{}\tiny  NCHW Pool 256x96x55x55 5x5 (cold caches)}}}%
      \put(3055,2090){\makebox(0,0)[l]{\strut{}\tiny RC: 0.34\%}}%
      \csname LTb\endcsname%
      \put(850,3593){\makebox(0,0)[l]{\strut{}\tiny Attainable RC: 10.99 \%}}%
      \colorrgb{0.00,0.00,0.00}%
      \put(3859,453){\rotatebox{90}{\makebox(0,0)[l]{\strut{}\tiny  NCHW Pool 256x96x55x55 5x5 (warm caches)}}}%
      \put(3074,2173){\makebox(0,0)[l]{\strut{}\tiny RC: 0.34\%}}%
      \put(4008,453){\rotatebox{90}{\makebox(0,0)[l]{\strut{}\tiny  NCHW16C Pool Avg 256x96x55x55 5x5 (cold caches)}}}%
      \put(4173,3397){\makebox(0,0)[l]{\strut{}\tiny RC: 6.85\%}}%
      \put(4096,453){\rotatebox{90}{\makebox(0,0)[l]{\strut{}\tiny  NCHW16C Pool Avg 256x96x55x55 5x5 (warm caches)}}}%
      \put(4179,3324){\makebox(0,0)[l]{\strut{}\tiny RC: 6.79\%}}%
    }%
    \gplgaddtomacro\gplfronttext{%
      \csname LTb\endcsname%
      \put(64,2841){\rotatebox{-270}{\makebox(0,0){\strut{}\tiny Atteinable GFLOPS/s}}}%
      \put(4446,56){\makebox(0,0){\strut{}\tiny Arithmetic Intensity [FLOPS/Byte]}}%
      \put(4446,5547){\makebox(0,0){\strut{}\tiny Intel(R) Xeon(R) Gold 6248 CPU @ 2.50GHz; (single socket)}}%
      \csname LTb\endcsname%
      \put(7958,5324){\makebox(0,0)[r]{\strut{}\small Roofline}}%
    }%
    \gplbacktext
    \put(0,0){\includegraphics{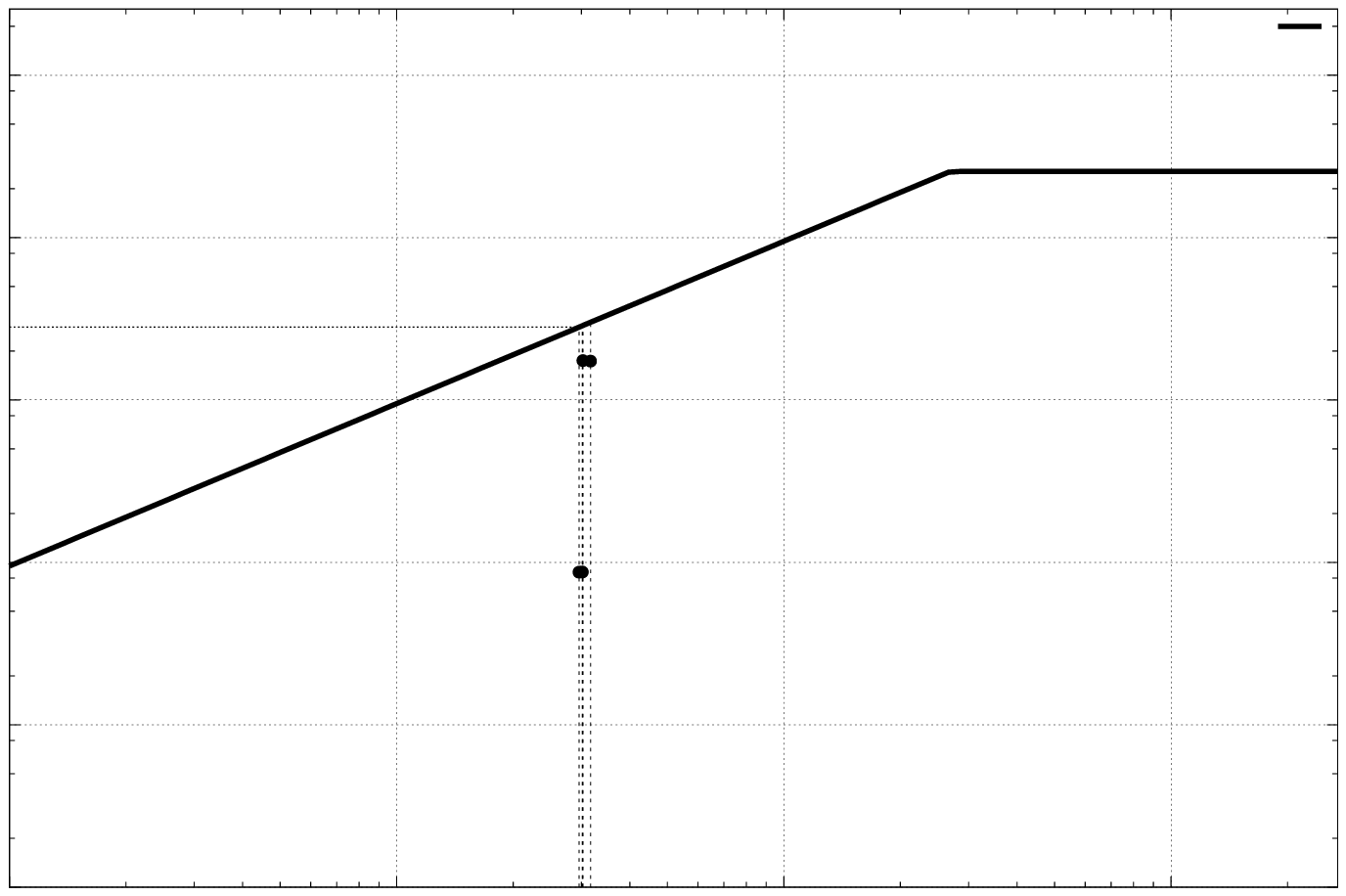}}%
    \gplfronttext
  \end{picture}%
\endgroup

\begingroup
  \fontfamily{Helvetica}%
  \selectfont
  \makeatletter
  \providecommand\color[2][]{%
    \GenericError{(gnuplot) \space\space\space\@spaces}{%
      Package color not loaded in conjunction with
      terminal option `colourtext'%
    }{See the gnuplot documentation for explanation.%
    }{Either use 'blacktext' in gnuplot or load the package
      color.sty in LaTeX.}%
    \renewcommand\color[2][]{}%
  }%
  \providecommand\includegraphics[2][]{%
    \GenericError{(gnuplot) \space\space\space\@spaces}{%
      Package graphicx or graphics not loaded%
    }{See the gnuplot documentation for explanation.%
    }{The gnuplot epslatex terminal needs graphicx.sty or graphics.sty.}%
    \renewcommand\includegraphics[2][]{}%
  }%
  \providecommand\rotatebox[2]{#2}%
  \@ifundefined{ifGPcolor}{%
    \newif\ifGPcolor
    \GPcolorfalse
  }{}%
  \@ifundefined{ifGPblacktext}{%
    \newif\ifGPblacktext
    \GPblacktexttrue
  }{}%
  \let\gplgaddtomacro\g@addto@macro
  \gdef\gplbacktext{}%
  \gdef\gplfronttext{}%
  \makeatother
  \ifGPblacktext
    \def\colorrgb#1{}%
    \def\colorgray#1{}%
  \else
    \ifGPcolor
      \def\colorrgb#1{\color[rgb]{#1}}%
      \def\colorgray#1{\color[gray]{#1}}%
      \expandafter\def\csname LTw\endcsname{\color{white}}%
      \expandafter\def\csname LTb\endcsname{\color{black}}%
      \expandafter\def\csname LTa\endcsname{\color{black}}%
      \expandafter\def\csname LT0\endcsname{\color[rgb]{1,0,0}}%
      \expandafter\def\csname LT1\endcsname{\color[rgb]{0,1,0}}%
      \expandafter\def\csname LT2\endcsname{\color[rgb]{0,0,1}}%
      \expandafter\def\csname LT3\endcsname{\color[rgb]{1,0,1}}%
      \expandafter\def\csname LT4\endcsname{\color[rgb]{0,1,1}}%
      \expandafter\def\csname LT5\endcsname{\color[rgb]{1,1,0}}%
      \expandafter\def\csname LT6\endcsname{\color[rgb]{0,0,0}}%
      \expandafter\def\csname LT7\endcsname{\color[rgb]{1,0.3,0}}%
      \expandafter\def\csname LT8\endcsname{\color[rgb]{0.5,0.5,0.5}}%
    \else
      \def\colorrgb#1{\color{black}}%
      \def\colorgray#1{\color[gray]{#1}}%
      \expandafter\def\csname LTw\endcsname{\color{white}}%
      \expandafter\def\csname LTb\endcsname{\color{black}}%
      \expandafter\def\csname LTa\endcsname{\color{black}}%
      \expandafter\def\csname LT0\endcsname{\color{black}}%
      \expandafter\def\csname LT1\endcsname{\color{black}}%
      \expandafter\def\csname LT2\endcsname{\color{black}}%
      \expandafter\def\csname LT3\endcsname{\color{black}}%
      \expandafter\def\csname LT4\endcsname{\color{black}}%
      \expandafter\def\csname LT5\endcsname{\color{black}}%
      \expandafter\def\csname LT6\endcsname{\color{black}}%
      \expandafter\def\csname LT7\endcsname{\color{black}}%
      \expandafter\def\csname LT8\endcsname{\color{black}}%
    \fi
  \fi
    \setlength{\unitlength}{0.0500bp}%
    \ifx\gptboxheight\undefined%
      \newlength{\gptboxheight}%
      \newlength{\gptboxwidth}%
      \newsavebox{\gptboxtext}%
    \fi%
    \setlength{\fboxrule}{0.5pt}%
    \setlength{\fboxsep}{1pt}%
\begin{picture}(8502.00,5668.00)%
    \gplgaddtomacro\gplbacktext{%
      \csname LTb\endcsname%
      \put(488,256){\makebox(0,0)[r]{\strut{}\tiny $0.1$}}%
      \csname LTb\endcsname%
      \put(488,1164){\makebox(0,0)[r]{\strut{}\tiny $1$}}%
      \csname LTb\endcsname%
      \put(488,2073){\makebox(0,0)[r]{\strut{}\tiny $10$}}%
      \csname LTb\endcsname%
      \put(488,2981){\makebox(0,0)[r]{\strut{}\tiny $100$}}%
      \csname LTb\endcsname%
      \put(488,3890){\makebox(0,0)[r]{\strut{}\tiny $1000$}}%
      \csname LTb\endcsname%
      \put(488,4798){\makebox(0,0)[r]{\strut{}\tiny $10000$}}%
      \csname LTb\endcsname%
      \put(536,176){\makebox(0,0){\strut{}\tiny $0.1$}}%
      \csname LTb\endcsname%
      \put(2828,176){\makebox(0,0){\strut{}\tiny $1$}}%
      \csname LTb\endcsname%
      \put(5119,176){\makebox(0,0){\strut{}\tiny $10$}}%
      \csname LTb\endcsname%
      \put(7411,176){\makebox(0,0){\strut{}\tiny $100$}}%
      \colorrgb{0.00,0.00,0.00}%
      \put(6065,4590){\makebox(0,0)[l]{\strut{}\tiny compute bound (Peak Runtime Compute: 100\%)}}%
      \put(3770,453){\rotatebox{90}{\makebox(0,0)[l]{\strut{}\tiny  NCHW Pool 256x96x55x55 5x5 (cold caches)}}}%
      \put(3080,2207){\makebox(0,0)[l]{\strut{}\tiny RC: 0.25\%}}%
      \csname LTb\endcsname%
      \put(850,3697){\makebox(0,0)[l]{\strut{}\tiny Attainable RC: 11.32 \%}}%
      \colorrgb{0.00,0.00,0.00}%
      \put(3840,453){\rotatebox{90}{\makebox(0,0)[l]{\strut{}\tiny  NCHW Pool 256x96x55x55 5x5 (warm caches)}}}%
      \put(3081,2144){\makebox(0,0)[l]{\strut{}\tiny RC: 0.22\%}}%
      \put(3997,453){\rotatebox{90}{\makebox(0,0)[l]{\strut{}\tiny  NCHW16C Pool Avg 256x96x55x55 5x5 (cold caches)}}}%
      \put(4142,3240){\makebox(0,0)[l]{\strut{}\tiny RC: 4.35\%}}%
      \put(4083,453){\rotatebox{90}{\makebox(0,0)[l]{\strut{}\tiny  NCHW16C Pool Avg 256x96x55x55 5x5 (warm caches)}}}%
      \put(4166,3444){\makebox(0,0)[l]{\strut{}\tiny RC: 6.57\%}}%
      \put(6801,357){\makebox(0,0)[l]{\strut{}\tiny RC - Runtime Compute}}%
    }%
    \gplgaddtomacro\gplfronttext{%
      \csname LTb\endcsname%
      \put(64,2841){\rotatebox{-270}{\makebox(0,0){\strut{}\tiny Atteinable GFLOPS/s}}}%
      \put(4446,56){\makebox(0,0){\strut{}\tiny Arithmetic Intensity [FLOPS/Byte]}}%
      \put(4446,5547){\makebox(0,0){\strut{}\tiny Intel(R) Xeon(R) Gold 6248 CPU @ 2.50GHz; (two sockets)}}%
      \csname LTb\endcsname%
      \put(7958,5324){\makebox(0,0)[r]{\strut{}\small Roofline}}%
    }%
    \gplbacktext
    \put(0,0){\includegraphics{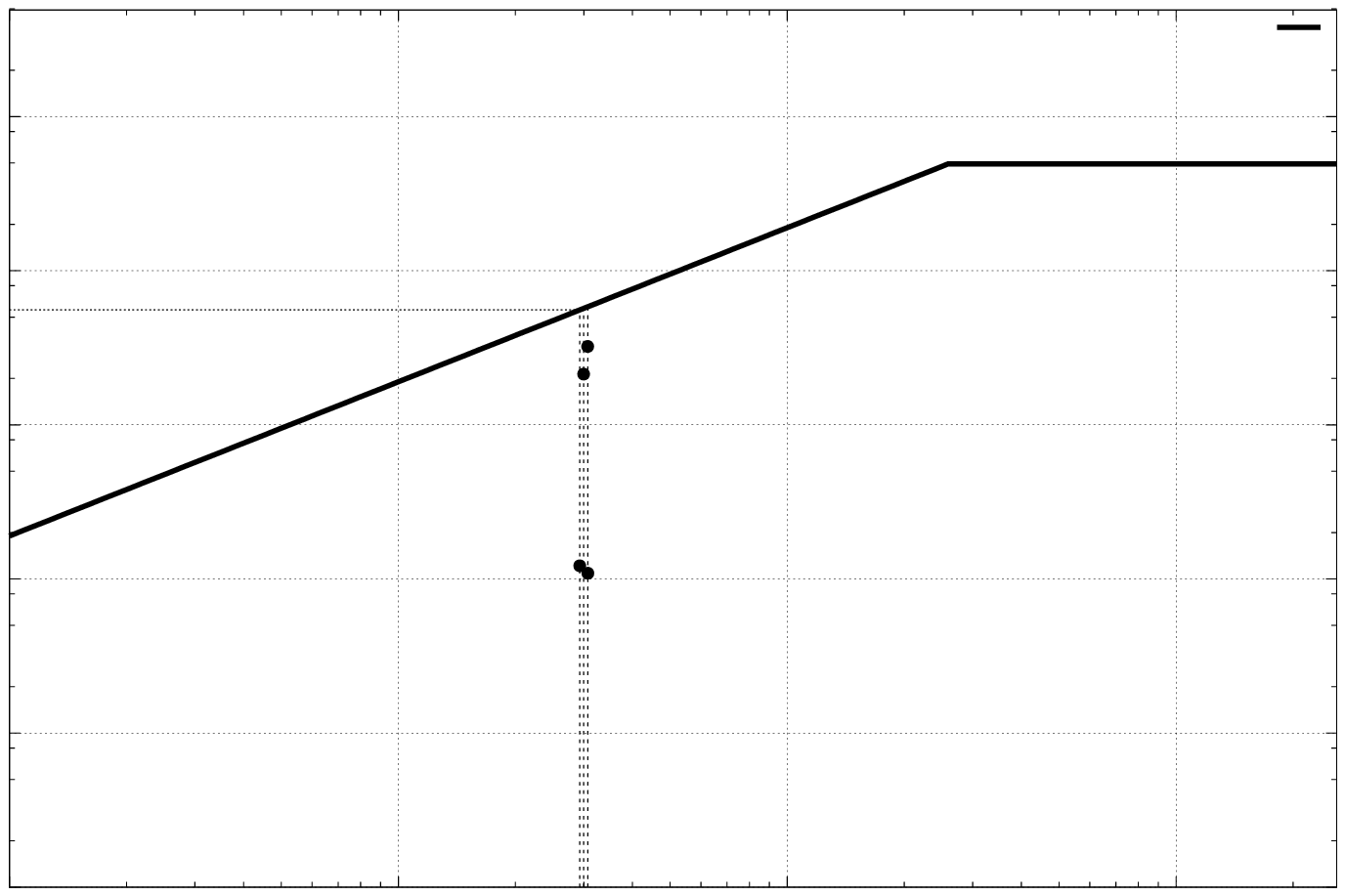}}%
    \gplfronttext
  \end{picture}%
\endgroup

\end{document}